\documentclass[]{aastex631}

\newcommand{\TDF}{NEP TDF}
\newcommand{\EBV}{\textsl{E}(\textsl{B}$-$\textsl{V})}%

\received{May 20, 2023}
\revised{August 5, 2023}
\accepted{August 28, 2023}

\shorttitle{NIR imaging in the NEP TDF}
\shortauthors{Willmer et al.}

\usepackage{graphicx}
\usepackage[english]{babel}
\usepackage{bibentry}
\usepackage{natbib}
\usepackage{tabularx}
\usepackage{blindtext}
\usepackage{tabularx}
\usepackage{multirow}
\usepackage{datetime}
\usepackage{parskip}
\newdateformat{mydate}{\THEDAY~ \monthname[\THEMONTH] \THEYEAR}
\hypersetup{urlcolor=blue}

\begin{document}

\title{PEARLS: Near Infrared Photometry in the JWST North Ecliptic Pole Time Domain Field\footnote{Based on observations taken with the MMT, a joint
    facility operated by the University of Arizona and the Smithsonian
    Institution.}}

\correspondingauthor{Christopher N. A. Willmer}
\email{cnaw@arizona.edu}

\author[0000-0001-9262-9997]{Christopher N. A. Willmer}
\affiliation{Steward Observatory, University of Arizona, 933 North
  Cherry Avenue, Tucson, AZ 85721, USA} 

\author[0000-0002-4245-2318]{Chun Ly}
\affiliation{Steward Observatory, University of Arizona, 933 North
  Cherry Avenue, Tucson, AZ 85721, USA} 

\author[0000-0003-3214-9128]{Satoshi Kikuta}
\affiliation{National Astronomical Observatory of Japan, 2-21-1 Osawa, Mitaka, Tokyo 181-8588, Japan} 

\author{S. A. Kattner}
\affiliation{MMT Observatory, University of Arizona
933 North Cherry Avenue Tucson, AZ 85721, USA}

\author[0000-0003-1268-5230]{Rolf A. Jansen}
\affiliation{School of Earth \& Space Exploration, Arizona State University, 550 E. Tyler Mall, Tempe, AZ 85287-1404, USA}

\author[0000-0003-3329-1337]{Seth H. Cohen}
\affiliation{School of Earth \& Space Exploration,
Arizona State University, 550 E. Tyler Mall, Tempe, AZ 85287-1404, USA}

\author[0000-0001-8156-6281]{Rogier A. Windhorst}
\affiliation{School of Earth \& Space Exploration,
Arizona State University, 550 E. Tyler Mall, Tempe, AZ 85287-1404, USA}

\author[0000-0003-3037-257X]{Ian Smail}
\affiliation{Centre for Extragalactic Astronomy, Department of Physics, Durham University, South Road, Durham, DH1 3LE, UK} 

\author[0000-0001-9052-9837]{Scott Tompkins}
\affiliation{School of Earth \& Space Exploration,
Arizona State University, 550 E. Tyler Mall, Tempe, AZ 85287-1404, USA}

\author[0000-0002-0005-2631]{John F. Beacom}
\affiliation{Center for Cosmology and AstroParticle Physics, 191 W. Woodruff Avenue, Columbus, OH 43210, USA}
\affiliation{Department of Physics, Ohio State University, 191 W. Woodruff Avenue, Columbus, OH 43210, USA}
\affiliation{Department of Astronomy, Ohio State University, 140 West 18$^{th}$ Avenue, Columbus, OH 43210, USA}

\author[0000-0003-0202-0534]{Cheng Cheng}
\affiliation{Chinese Academy of Sciences, South America Center for Astronomy, National Astronomical Observatories, CAS, Beijing 100101, China}

\author[0000-0003-1949-7638]{Christopher J. Conselice}
\affiliation{Jodrell Bank Centre for Astrophysics, University of Manchester, Oxford Road, Manchester M13 9PL, UK}

\author[0000-0003-1625-8009]{Brenda L. Frye}
\affiliation{Steward Observatory, University of Arizona, 933 North
  Cherry Avenue, Tucson, AZ 85721, USA} 

\author[0000-0002-6610-2048]{Anton M. Koekemoer}
\affiliation{Space Telescope Science Institute, 3700 San Martin Drive
Baltimore, MD 21218, USA}
\author[0000-0001-6145-5090] {Nimish Hathi}
\affiliation{Space Telescope Science Institute, 3700 San Martin Drive Baltimore, MD 21218, USA}
\author[0000-0003-4738-4251]{Minhee Hyun}
\affiliation{Korea Astronomy and Space Science Institute, 776
  Daedeok-daero, Yuseong-gu, Daejeon 34055, Republic of Korea}

\author[0000-0002-8537-6714]{Myungshin Im}
\affiliation{SNU, Astronomy Research Center, Astronomy program,
  Dept. of Physics \& Astronomy, Seoul National University, 1 Gwanak-ro, 
  Gwanak-gu, Seoul 08826, Republic of Korea}
  
\author[0000-0002-9895-5758]{S.\ P.\ Willner}
\affiliation{Center for Astrophysics \textbar\ Harvard \& Smithsonian, 60 Garden Street, Cambridge, MA, 02138, USA}
  
\author[0000-0002-7791-3671] {X. Zhao}
\affiliation{Center for Astrophysics \textbar\ Harvard \& Smithsonian, 60 Garden Street, Cambridge, MA, 02138, USA}

\author{Walter A. Brisken}
\affiliation{Long Baseline Observatory, 1003 Lopezville Rd, Socorro, NM 87801, USA}

\author{F. Civano}
\affiliation{NASA Goddard Space Flight Center, Greenbelt, MD 20771, USA}

\author[0000-0001-7363-6489]{William Cotton}
\affiliation{National Radio Astronomy Observatory, 520 Edgemont Road,
  Charlottesville, VA 22903, USA} 

\author[0000-0002-0797-0646]{G\"unther Hasinger}
\affiliation{ESA, European Space Astronomy Centre (ESAC),Camino bajo del Castillo, s/n, Urbanizaci\'on Villafranca del Castillo, Villanueva de la Ca\~nada, 
E-28692 Madrid, Spain}

\author[0000-0002-2203-7889]{W. Peter Maksym}
\affiliation{Center for Astrophysics \textbar\ Harvard \& Smithsonian, 60 Garden Street, Cambridge, MA, 02138, USA}

\author[0000-0002-7893-6170]{Marcia J. Rieke}
\affiliation{Steward Observatory, University of Arizona, 933 North
  Cherry Avenue, Tucson, AZ 85721, USA} 

\author[0000-0001-9440-8872]{Norman A. Grogin} 
\affiliation{Space Telescope Science Institute, 3700 San Martin Drive Baltimore, MD 21218, USA}

\begin{abstract}
  We present Near-Infrared (NIR) ground-based \textsl{Y}, \textsl{J}, \textsl{H}, and \textsl{K} imaging obtained in the James Webb Space Telescope North Ecliptic Pole Time Domain Field (\TDF) using the MMT-Magellan Infrared Imager and Spectrometer (MMIRS) on the MMT.
  These new observations cover a field of approximately 230 arcmin$^2$ in \textsl{Y}, \textsl{H}, and \textsl{K} and 313 arcmin$^2$ in \textsl{J}. Using Monte Carlo simulations we estimate a 1$\sigma$ depth relative to the background sky of (\textsl{Y, J, H, K}) = (23.80, 23.53, 23.13, 23.28) in AB magnitudes for point sources at a 95\% completeness level.

  These observations are part of the ground-based effort to characterize this region of the sky, supplementing space-based data obtained with Chandra, NuSTAR, XMM, AstroSat, HST, and JWST. This paper describes the observations and reduction of the NIR imaging and combines these NIR data with archival imaging in the visible, obtained with the Subaru Hyper-Suprime-Cam, to produce a merged catalog of 57,501 sources. 
  The new observations reported here, plus the corresponding multi-wavelength catalog, will provide a baseline for time-domain studies of bright sources in the {\TDF}.
\end{abstract}


\section{Introduction} \label{sec:intro}
One of the greatest discoveries in the last few decades is the existence of Dark Energy, which was inferred from the observations of Type Ia supernovae ranging from the local universe to redshifts of 1 and above. This discovery has prompted several large scale surveys of galaxies, e.g., SDSS-IV \citep{Albareti2017}, Palomar Transient Factory \citep{Rau2009}, J-PAS \citep{Bonoli2021}, and DESI \citep{Dey2019} among others, to better characterize Dark Energy and understand its origin. Because the detection and follow-up of very distant supernovae from the ground becomes very challenging at redshifts of $z\sim2$ and above, the use of space-based facilities becomes essential. However, a major limitation in using the space observatories is the ability to monitor regions of the sky on a somewhat regular time cadence, that allows identifying and following the changes transient sources undergo.
 
The combination of  mirror size,  sensitivity, and smaller Galactic attenuation in the infrared, allow the James Webb Space Telescope (JWST) to reach much fainter flux levels than ever before, and make it an ideal instrument to detect sources presenting changes, be it through proper motions (e.g., outer Solar System bodies and nearby stellar and sub-stellar objects), or in light output (e.g., supernovae, variable stars, AGN).
\citet{Jansen2018} proposed taking advantage of a region of the sky that can be observed by JWST any time of the year to detect transient, variable, and moving objects much fainter than is feasible from the ground even with near-future facilities like the Rubin Observatory.
Examining the distribution of stars and mid-infrared emitters (bright stars, galactic cirrus), \citet{Jansen2018} selected a patch of sky close to the North Ecliptic Pole (NEP) within JWST's northern continuous viewing zone (CVZ), which is a position on the sky used for spacecraft house keeping and will be frequently visited by the telescope. The position chosen in the northern CVZ is centered at (R.A., Dec.) J2000 = (17:22:47.896, +65:49:21.54), and has a particularly low number density of bright ($m_K$ $\leq$ 15.5 mag) sources and low Galactic extinction (\EBV\ $\lesssim$ 0.03 mag) where such time-monitoring survey work will be feasible and efficient.
This area was dubbed the JWST NEP Time Domain Field (\TDF) and will be a prime field for time-domain measurements during the lifetime of the telescope. It is also one of two CVZ targets of the Prime Extragalactic Areas for Reionization and Lensing Science (\textsl{PEARLS}) JWST/GTO program  2738 (PIs: Windhorst \& Hammel) \citep{Windhorst2023}, the other being the IRAC Dark Field \citep{Yan2023} a few degrees away.

Prior to 2016, the \TDF\ was covered by imaging data coming the Sloan Digital Sky Survey \citep{Alam2015} and Pan-STARRS \citep{Tonry2012} in the visible, 2MASS \citep{Skrutskie2006} in the near-infrared (NIR), and WISE \citep{wright2010} in the mid-infrared.
While extremely valuable, these data were not deep enough to detect the very faint sources necessary to vet the astrometry and photometry of the JWST \TDF\ observations.  This prompted a multi-observatory effort to collect visible and NIR data for the calibrations and to provide the first epoch observations to identify bright transients and variable sources.
These visible and NIR data also provide counterparts for observations obtained at other wavelengths, both from space, e.g., Chandra (PI: W.~Maksym),  NuSTAR (PI: F.~Civano; Zhao et al. \citeyear{Zhao2021}), XMM (PI: F.~Civano), eROSITA (PI: A.~Merloni \& R.~Sunyaev),  AstroSat (PI: K.~Saha), and HST (PIs: R.~Jansen \& N.~Grogin) and from the ground: LBT/LBC (PI: R.~Jansen), J-NEP (PI: S.~Bonoli \& R. Dupke; \citealt{JNEP}), HiPeRCAM/Gran Telescopio de Canarias (PI: V.~Dhillon), Hyper-Suprime-Cam/Subaru (HEROES, PI: G.~Hasinger \& E.~Hu; \citealt{Songaila2018}; \citealt{Taylor2023}), NOEMA (PI: S.~Cohen), SCUBA2 (PIs: M.~Im \& I.~Smail; \citealt{Hyun2023}), SMA (PI: G.~Fazio), VLA (PI: R.~Windhorst; \citealp{Hyun2023}; S.~Willner, 2023, \apj ~submitted), and LOFAR (PI: R.~van\,Weeren). 

In this paper, we report the results of a NIR imaging survey in the \TDF\null. These observations increase the depth relative to 2MASS by more than 5 magnitudes and provide baseline observations for brighter sources detected by JWST\null.  The NIR observations cover a wider field than targeted by JWST and provide NIR-counterparts for sources outside the  JWST footprint but which are contained within the fields-of-view of the other space observatories. We supplement the NIR infrared catalogs with visible data obtained from our own reduction of archival Subaru Hyper-Suprime-Cam (HSC) data taken in this part of the sky.
In \S~2, we describe the observations, in \S~3, the reduction and calibrations, in \S~4, the analyses of MMIRS and the archival HSC data, and in \S~5, our conclusions. 
All magnitudes are quoted in the AB system \citep{Oke_Gunn_1983} and, following \citet{2018ApJS..236...47W}, we adopt the Vega to AB transformations (\textsl{MMIRS-Y}, \textsl{MMIRS-J}, \textsl{MMIRS-H}, \textsl{MMIRS-K}, \textsl{MMIRS-K$_{\rm spec}$}) = (0.574, 0.891,1.333, 1.836, 1.840) mag. These filters will be noted throughout the text as \textsl{Y, J, H, K} and \textsl{K$_{\rm spec}$}.

\begin{figure*}[htb!]
\plotone{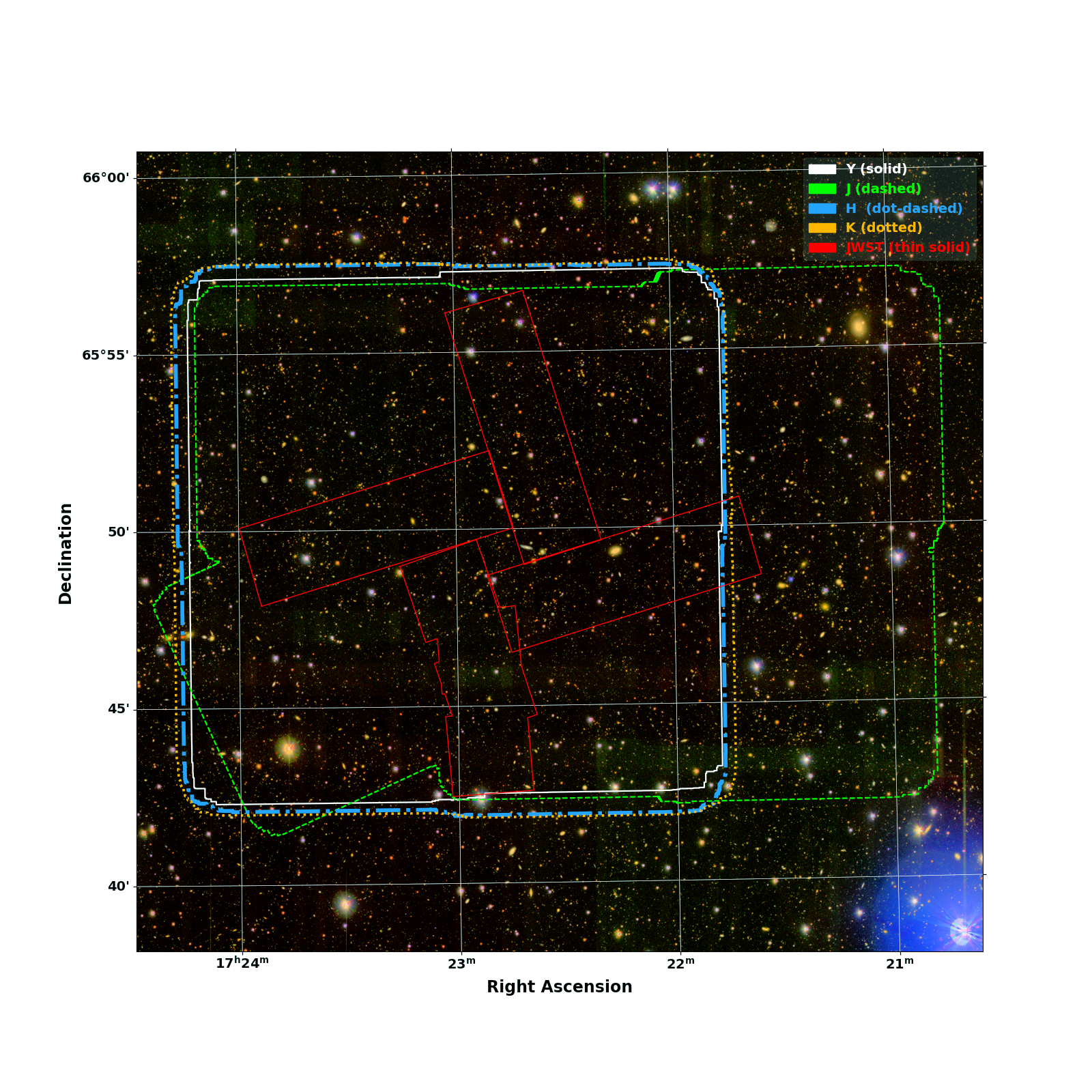}
\caption{\label{fig:nep_regions}
Footprint of the imaging for different MMIRS filters in the \TDF\ using a
$\sim$23\farcm9$\times$22\farcm6 Subaru HSC 
\textsl{g, i2, z} color composite as backdrop, with North up and East to the left with the grid showing coordinates in  sexagesimal units. The \TDF\ is centered at (R.A., Decl.)J$_{2000}$ = (17:22:47.896, +65:49:21.54). The white solid line outlines the \textsl{Y} imaging, the green dashed line outlines \textsl{J}, the thick blue dash-dotted line \textsl{H}, and the orange dots \textsl{K}. Most outlines overlap each other, with the exception of \textsl{J}.  The \textsl{K} outline combines observations taken with the (short) \textsl{K} and \textsl{K$_{\rm spec}$} filters; the latter occupy the SE corner quadrant. The figure also includes the outline of the completed JWST/NIRCam observations as a thin red solid line. When the MMIRS observations were initiated, a preliminary \TDF\ pointing and set of NIRCam position angles was used. The NIRCam observations as executed use a field center offset to the W, such that the full \TDF\ is covered with MMIRS only in \textsl{J}.}
\end{figure*}

\section{Observations} \label{sec:observations}
We imaged the JWST \TDF\ with the MMT-Magellan Infrared Imager and Spectrometer (MMIRS; McLeod et al. \citeyear{McLeod2012}, Chilingarian et al.\ \citeyear{Chilingarian2015}) on the MMT. The  observations were taken in queue-mode over 31 nights between UT 2017 May 14 and UT 2019 June 21. In addition to operating MMIRS, the queue observers also recorded the sky conditions, overheads, and any anomalies for each executed group of exposures (``blocks''). A log of these observations is provided in Table~\ref{tab:log} (Appendix~A). 
MMIRS is equipped with a single HAWAII-2RG (H2RG) detector with 2040\,$\times$\,2040 light-sensitive pixels\footnote{The remaining pixels (``reference pixels'') are used to monitor the detector thermal variations.}. 
The detector pixel scale of $\sim$0\farcs21 pixel$^{-1}$ gives a field of view of $\sim$6\farcm8$\times$6\farcm8, so that 4 pointings (a 2$\times$2 mosaic) are required to cover the 14$'$ diameter \TDF\ footprint.
MMIRS allows using different gain values --- the default low gain of 2.68~e$^-$/DN and a high gain of 0.95~e$^-$/DN, designed for the observations of faint sources. 
The detector is read out non-destructively using 32 amplifiers, generating ``ramps'' for each exposure (``data cubes'', where the third dimension contains the individual readouts).
The observations used the \textsl{Y}, \textsl{J}, \textsl{H}, and \textsl{K} filters; however, by accident, during the instrument setup early in the survey, some observations were taken using the wider spectroscopic filter \textsl{K$_{\rm spec}$}, others with the (short) \textsl{K} using the high gain value of 0.95~e$^-$/DN rather than the default 2.68~e$^-$/DN. Because of the different throughputs of the \textsl{K} and \textsl{K$_{\rm spec}$} filters and the different noise characteristics due to detector gain (which are discussed in greater detail in Section~\ref{sec:kdiff}), we differentiate  the observations according to filter and gain values;  thus, the (short) \textsl{K} observations  are represented as \textsl{K$_{2.68}$} and \textsl{K$_{0.95}$}.
To keep the sky background and source flux within the roughly linear regime and minimize the number of saturated sources, the integration time of individual exposures ranged from $\sim$8~s in \textsl{K} to$\sim$ 57~s in \textsl{Y}.
The ramps for the \TDF\ observations used the \textsf{LogGain/ramp\_4.5s} readout mode for \textsl{Y} and \textsl{J} and \textsf{ramp\_1.475} for \textsl{H} and \textsl{K/K$_{\rm spec}$}, generating data cubes with 14, 12, (8, 11)\footnote{Due to improvements on the detector read out implemented in 2019, the exposure times for \textsl{H} observations were increased from $\sim$11 to 16\,s, improving the observation efficiency.}, and 7 frames in (\textsl{Y}, \textsl{J}, \textsl{H}, \textsl{K+K$_{\rm spec}$}).
For each science image, a corresponding sky-camera exposure was stored, which allows monitoring changes in the sky conditions over the duration of each observing block. At the end of each night, a series of dark frames were taken with integration times identical to those used for the science exposures.

After each exposure, the telescope was moved using a pseudo-random dither pattern
to a new position at least 3 times the expected FWHM (assumed as 0\farcs8) away in order to mitigate both bad pixels and persistence.
Table~\ref{tab:summary} shows the exposure times per dither, the total number of images in each filter combining all quadrants and the average exposure time per pixel. The accumulated total on-target science exposure time was $\sim$3.9, 7.1, 5.9, and 9.7\,hours in \textsl{Y}, \textsl{J}, \textsl{H}, and \textsl{K/K$_{\rm spec}$}, respectively.

Figure~\ref{fig:nep_regions} shows outlines of the MMIRS NIR observations reported here, as well as outlines of the JWST observations in the \TDF\ which are now complete, overlaid on a background 
\textsl{g}, \textsl{i2}, \textsl{z} color composite constructed from our reduction of archival HSC observations taken in 2017 for the HEROES survey (\citealp{Songaila2018}; \citealp{Taylor2023}).


\begin{deluxetable*}{cclccccl}[htb!]
\tabletypesize{\scriptsize}
\tablecaption{Summary of MMT/MMIRS Observations of the \TDF\label{tab:summary}}
\tablewidth{0pt}
\tablehead{
\colhead{Filter} & \colhead{Gain} & \colhead{ZP$_{AB}$} &
\colhead{Exposure} &\colhead{No. Exposures} & \colhead{Total} &\colhead{Exposure} &\colhead{Fields}\\
{} & {} & {} & per image & {} & exposure & per pixel & {}\\
\colhead {} & \colhead{e$^-$DN$^{-1}$} & \colhead{}&
\colhead{s}&\colhead{}& \colhead{s} &\colhead{s} & \colhead{}
}
\startdata
\textsl{Y} & 2.68 & 21.30$\pm$0.01  & 57.536 &  245 & 14096 & 3558 & NE, ~NW, ~SE, ~SW\\
\textsl{J} & 2.68 & 21.46$\pm$0.01  & 48.684 &  528 & 25705 & 3535 & NE, ~NW, ~SE, ~SW, ~center, ~NW1, ~SW1\\
\textsl{H} & 2.68 & 23.33$\pm$0.01  & 11.802, 16.228 & 693\tablenotemark{a}, 809\tablenotemark{b} & 21307 & 5237& NE\tablenotemark{a,b}, ~NW\tablenotemark{a}, ~SE\tablenotemark{b}, ~SW\tablenotemark{b}\\
\textsl{K$_{2.68}$} & 2.68 & 23.37$\pm$0.01  &  8.852 & 2584 & 22874 & 823 & NE, ~NW, ~SW\\
\textsl{K$_{0.95}$} & 0.95 & 24.76$\pm$0.01  &  8.852 &  580 & 5134 & 4755 & SW, ~SE\\
\textsl{K$_{\rm spec}$}&2.68& 23.78$\pm$0.04 &  8.852 &  801 & 7090 & 3848 & SE\\
\textsl{K+K$_{\rm spec}$} &{...} & 23.58$\pm$0.04 & {...}& 3735 & 35098 & 7983 & Mosaic combining all K imaging\\
\enddata
\tablenotetext{a}{11.802 s exposures}
\tablenotetext{b}{16.228 s exposures}
\end{deluxetable*}


\clearpage
\section{Reduction} \label{sec:reduction}

\subsection{Individual images}
The reduction of the MMIRS imaging followed a similar procedure as
used by \citet{Labbe2003}, \citet{McCracken2012}, and \citet{Pello2018},
taking into account some of the characteristics of the H2RG detector.
The first step in the reduction was subtracting the dark current from
each science image plane using the average of the corresponding plane in the
set of  dark images. Next, image counts were 
corrected for non-linearity using look-up tables compiled by the MMIRS
team, cosmic rays were removed and cross-talk subtracted, after which the
count-rate images (``slopes'') for each ramp in units of data-numbers per second ($DN/s$) were calculated.
These reduction steps were carried out using the \citet{Chilingarian2015} \texttt{IDL} procedures\footnote{{\url{https://bitbucket.org/chil\_sai/mmirs-pipeline/wiki/Home}}} with adaptations for use with imaging data by one of the authors (CNAW) which are briefly described in Appendix~B\null. The procedures as modified are publicly available at
{\url{https://github.com/cnaw/mmirs\_imaging}}.

The sky subtraction followed the \texttt{IRAF} script {\bf{xdimsum}}
\citep{xdimsum}, which was specifically designed for the reduction of
near-infrared data; similar algorithms were used by \citet{Thompson1999} and \citet{Huang2003}. The first step in the process was removing a
baseline sky value from each image and storing the reciprocal of this
value.\footnote{\citet{xdimsum} used a median with iterations to remove outliers; here we used the outlier-resistant bi-weight estimator of \citet{Beers1990}.}
The initial sky value per slope image was obtained taking a
pixel-by-pixel average using the 16 (baseline-subtracted) images taken nearest in time
(both before and after a given exposure) with no masking of sources
or bad pixels.  In the case of images at 
the beginning or end of a sequence, we used the nearest images in time
following or preceding a given observation (still totalling 16). 
The next step calculated the ratio of the image divided by the sky,
from which the median ratio (after outlier rejection) was obtained.
The sky image multiplied by this median ratio was then subtracted from
the slope image.

Initial source catalogs per (sky-subtracted) image were calculated
using \texttt{SExtractor} 
\citep{BertinArnouts}, and masks were created using the segmentation
images, where each detection has its segmentation map extended radially by 2
pixels (0\farcs42) to mask any residual source 
light close to the detection-level isophote. The initial catalogs were 
also used to identify ``ghost'' sources due to image persistence.
This was done by examining the brightest sources and using
the pixel coordinates between sequential images; the persistence
images must fall $\lesssim$8 pixels away from the pixel position in the
previous exposure occupied by the bright star and be fainter by more
that 2 magnitudes. In general, after  
two ramps the remaining signal is negligible compared to the sky
background, though in the case of very bright stars, some residuals
can be detected even after $\sim$200~s (in the case of the
\textsl{Y} exposures, which are the longest integrations we used). All
persistence ghosts had their segmentation maps expanded radially by 5 pixels.
Bad pixels (dead pixels, hot pixels) were also flagged and their
positions noted in the segmentation images and added to the image
masks for each slope image. 

After the initial sky subtraction step, we measured the seeing of
individual images with the aim of optimizing the mosaic depth relative to the
image resolution.
To estimate the seeing, we used the sky-subtracted images to identify stellar sources in each image using 
distributions of the instrumental surface brightness of the brightest pixel (\texttt{$\mu_{max}$}) versus the detected area (\texttt{isoarea\_image}) to remove hot pixels and residual cosmic rays, and the half-light radius (\textsf{flux\_radius}) versus the ``total'' magnitude (\textsf{mag\_auto}) all calculated by \texttt{SExtractor}. The total magnitudes used a minimum Kron radius of 3.5 pixels (1\farcs47 diameter) and a Kron factor of 2.5.
The average FWHM for each image was  calculated from the measurements of these candidate stars using the bi-weight estimator \citep{Beers1990} .

\begin{figure*}[htb!]
\plottwo{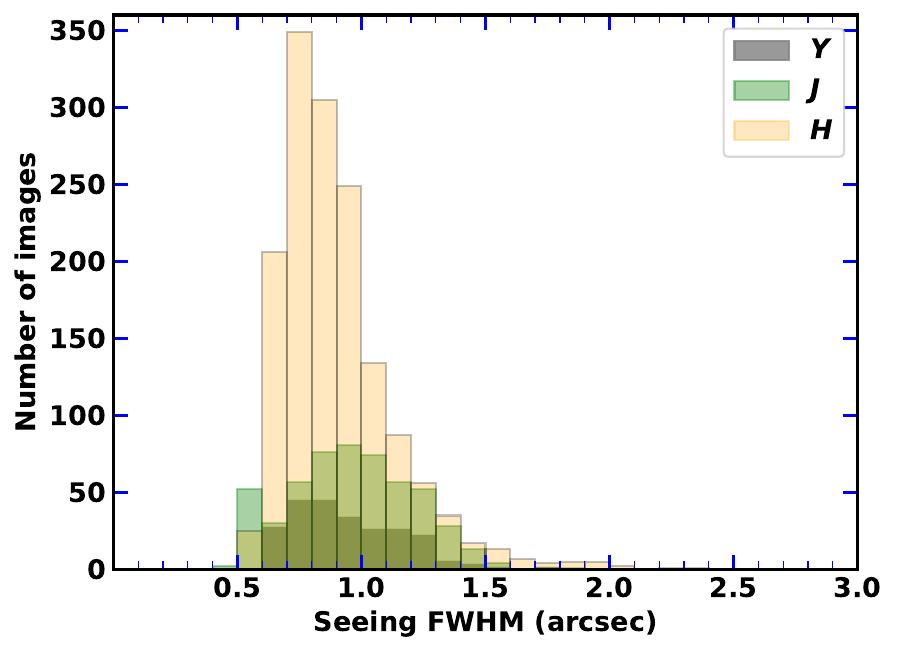}{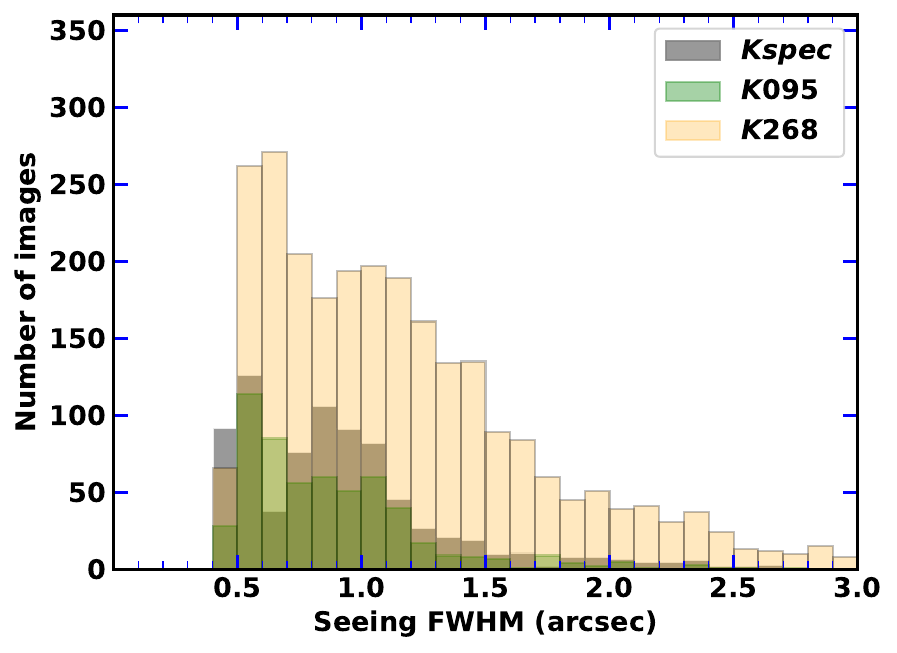}
\caption{Distribution of the seeing FWHM in arcsec measured for sources classified as stars in each individual image. The left panel shows the distributions for \textsl{Y}, \textsl{J} and \textsl{H}, and the right panel the distributions for the low and high gain \textsl{K$_{2.68}$, K$_{0.95}$} short \textsl{K} filter and
\textsl{K$_{\rm spec}$} as indicated in the key. The median seeing values for the different configurations are (\textsl{Y, J, H, K$_{0.95}$, K$_{2.68}$, K$_{\rm spec}$}) = (0\farcs90, 0\farcs96, 0\farcs85, 0\farcs80, 1\farcs06, 0\farcs86).
The distribution of \textsl{K$_{2.68}$} values reflects the variation of sky conditions over the several nights the data were collected. When making mosaics for the whole field, a maximum seeing cut at 2\farcs1 was adopted as a compromise between attaining depth and maintaining resolution; this cut mainly affects the \textsl{K$_{2.68}$} measurements. \label{fig:seeing}}
\end{figure*}

Figure~\ref{fig:seeing} shows the number of images in each filter for a
given FWHM. 
The left panel shows the distribution for the \textsl{Y, J} and \textsl{H} filters; the right panel shows the distribution for the observations using MMIRS with the \textsl{K$_{\rm spec}$}, \textsl{K$_{0.95}$} and \textsl{K$_{2.68}$} configurations. 
The majority of observations peak under $\sim$1$''$ values, and the distribution of the \textsl{K$_{\rm spec}$} \textsl{K$_{0.95}$} and \textsl{K$_{2.68}$} reflect the varying conditions during the observations, taken over several nights between 2017 June and 2017 November (see Figure~\ref{fig:seeinglog} for the variations over individual nights). In the case of \textsl{K$_{0.95}$}, data were taken in a single night, for \textsl{K$_{\rm spec}$} over the course of two nights. Most of the \textsl{K} values are $\leq$1\farcs5 but have a somewhat
broad distribution, particularly in the case of the \textsl{K$_{2.68}$} observations. 

A second calculation of the sky background was then
performed using the object masks, after which we refined the
World Coordinate System (WCS) of individual images using the
{\url{astrometry.net}} software of \citet{Lang2010} 
and the GAIA DR3 catalogue \citep{GAIADR3} as reference. The
improved astrometry for individual images has typical RMS residuals $\sim$0\farcs03.
During this phase of sky-subtraction, weights for individual images, described by the inverse variance due to the total number of photons in an exposure, the readout noise (3.14 $\rm e^-$), and the gain, have also aggregated a contribution due to the seeing:
\begin{widetext}
\begin{equation}
  {\rm weight}_{i,j} =
  \frac{1}{({\rm count\_rate}_{i,j} \times {\rm exptime} + {\rm {readnoise/gain}}^2) \times {\rm FWHM}^2}.
\end{equation}
\end{widetext}
Masked pixels were assigned inverse-variance values of 0 so \texttt{swarp} \citep{Bertin2002}
ignores these pixels when assembling mosaics, while the lowest
weight values were set to 10$^{-9}$ for the \texttt{SExtractor} parameter \textsf{weight\_thresh}; pixels with associated weights lower than this value are ignored by  \texttt{SExtractor}.
With the image masks and weights in place, another round of object
detection was carried out. These new catalogs were used by \texttt{scamp}
\citep{Bertin2006} to include higher-order terms in the astrometry
providing WCS keywords compatible with \texttt{swarp}
\footnote{In particular, the Tan-SIP transform used
  by \citet{Lang2010} is not recognized by Astromatic software
  (\texttt{SExtractor}, \texttt{scamp}, and  \texttt{swarp}).}.
The astrometric calibration of \texttt{scamp} also
used the GAIA DR3 catalog matched to sources classified as stellar
on individual slope images.
The zero-point magnitude offsets between individual slope images as
well as initial ensemble zero-point offsets were calculated using sources matched to 2MASS \citep{Skrutskie2006} for \textsl{J}, \textsl{H}, \textsl{K} and 
\textsl{K$_{\rm spec}$} 
and Pan-STARRS \citep{Tonry2012} for  the \textsl{Y} band. These
corrections by \texttt{scamp} were written into ASCII headers subsequently
read by \texttt{swarp} when creating the full-field mosaics.

\subsection{Mosaics}
To create the full-field mosaics,
we followed a similar procedure as Ashcraft et al. (\citeyear{Ashcraft2018}), Ashcraft et al. (\citeyear{Ashcraft2023}) and \citet{McCabe2023} where images in a given filter are combined in two different ways, one to optimize the spatial resolution at the cost of depth, and the other to optimize the depth at the cost of resolution. In these works, the optimal resolution images result from stacking the best seeing images, comprising about 10\% of the total number of images \citep{Ashcraft2018}, while for the optimal depth the 5--10\% worst seeing images are excluded (\citealt{Ashcraft2018, McCabe2023}); typically the cutoff is made at $\sim$ 2\farcs0, the deeper mosaic reaching $\sim$ 1 magnitude fainter than the higher resolution mosaic \citep{McCabe2023}.

Because the aim of the NIR imaging was to provide a first epoch of variability studies, we opted to reach as faint as possible, and following these works, we cut at 2\farcs1 when stacking the images. The greatest impact of this cut is on the \textsl{K$_{2.68}$} images, where 213 out of 2571 ($\sim$ 8\%)  were excluded. These numbers are  smaller for the remaining setups---26/796 ($\sim$3\%) for \textsl{K$_{\rm spec}$}, 7/563 for \textsl{K$_{0.95}$}, 3/1502 for \textsl{H} and 1/248 for \textsl{Y}. No \textsl{J} images were excluded.

In addition to mosaics of individual images, stacked images combining the \textsl{Y+J},
\textsl{K+K$_{\rm spec}$}, and \textsl{H+K+K$_{\rm spec}$} were created to have deeper
(or in the case of \textsl{K+K$_{\rm spec}$}, full field) coverage. In addition
to the image mosaics, \texttt{swarp} calculated a stacked weight image
which includes the contribution from individual image weights. For
the output mosaics, we adopted the  pixel size of 0\farcs168 and
center position (17:22:48.1298, +65:50:14.853, J2000) used in the Subaru/HSC
imaging discussed in \S4.3. The final mosaics were combined using
\texttt{swarp} parameters \textsf{combine\_type=clipped} with \textsf{clip\_sigma=3.0}.

The source catalogs were generated in a three-step process. The
first step used \texttt{SExtractor} with a high detection threshold (5$\sigma$) per pixel relative to the average background, optimized to
detect stars; these candidate sources were then processed by
\texttt{PSFEX} \citep{Bertin2011} to calculate the average
point-spread function for each mosaic.
The PSFs were used in a second pass of \texttt{SExtractor} to
calculate the photometry and image classification using 
the parameters listed in Table~\ref{tab:se_params}. 
All runs used the dual-image mode, where the detection images were mosaics
combining \textsl{J+Y} for \textsl{J} and \textsl{Y},
\textsl{H+K+K$_{\rm spec}$} for \textsl{H}, and combined \textsl{K+K$_{\rm spec}$} for the
\textsl{K} and \textsl{K$_{\rm spec}$} images.
The positions of these  catalogs are on the GAIA-DR3 system, and using GAIA-DR3 stars we find that the MMIRS positions have uncertainties $\lesssim$ 0\farcs070.
The photometric zero-points reported in Table~\ref{tab:se_params} were 
calculated following
\citet{Almeida-Fernandes2022}, who used external photometric measurements coming from
wide-area surveys (e.g., GAIA, SDSS) with the stellar
population models of \citet{Coelho2014} to calculate the unknown
zero-point offsets of uncalibrated filters. This procedure creates a grid of
flux measurements by folding the stellar population models
with the filter curves of the reference data and considering several
values of Galactic extinction. 
By means of $\chi^2$ minimization, the best-fitting model is found that also provides an estimate of the Galactic (and atmospheric)
extinction. From the ensemble of fitted stars, the best fitting
zero-point is then estimated.

In this calibration, we matched the MMIRS-derived catalogs with 
GAIA, SDSS, Pan-STARRS and 2MASS and only used sources that were classified as
stars in GAIA and which had measurements in all other catalogs. While the number of sources used is small (ranging from 32 in the case of \textsl{K$_{\rm spec}$} to 136 for \textsl{J}), particularly compared to the thousands used by 
\citet{Almeida-Fernandes2022}, the average values for the zero-point estimates show uncertainties of only $\sim$1\%--4\%. These are much smaller than the estimates calculated using \texttt{scamp} which are of the order of 10\%--30\% (per image) when using 2MASS measurements alone. These final zero-point
values are shown in the third column of Table~\ref{tab:summary}.

\begin{deluxetable*}{lll}[htb!]
\tablecaption{MMIRS \texttt{SExtractor} detection parameters\label{tab:se_params}.}
\tablewidth{0pt}
\tablehead{
\colhead{Parameter} & \colhead{value} & \colhead{note}}
\startdata
DETECT\_THRESH    & 1.0 & $\sigma$ above mean background\\
ANALYSIS\_THRESH  & 1.0 & $\sigma$ above mean background\\
DETECT\_MINAREA   & 10  & pixels \\
DEBLEND\_NTHRESH  & 32  & {} \\
DEBLEND\_MINCONT  & 0.005 & {} \\
FILTER\_NAME      & tophat\_5.0\_5$\times$5.conv\\
PHOT\_FLUXFRAC    & 0.5   & {} \\
PHOT\_AUTOPARAMS  & 2.5,3.5 & Kron factor, minimum radius\\
PHOT\_AUTOAPERS   & 0.0,0.0 & {}\\
CLEAN\_PARAM      & 0.3   & {} \\
WEIGHT\_TYPE      & MAP\_WEIGHT & {} \\
RESCALE\_WEIGHTS  &  Y      & {} \\
WEIGHT\_THRESH    &  1.e-09 & {} \\
SATUR\_LEVEL      &   9.9   & DN/s  \textsl{Y}\\
ZERO\_POINT       & 21.2991 & mag$_{AB}$ \textsl{Y}\\
SATUR\_LEVEL      &   10.   & DN/s  \textsl{J}\\
ZERO\_POINT       & 21.4560 & mag$_{AB}$ \textsl{J}\\
SATUR\_LEVEL      &   68.   & DN/s \textsl{H}\\
ZERO\_POINT       & 23.3339 & mag$_{AB}$ \textsl{H}\\
SATUR\_LEVEL      &  350.0  & DN/s \textsl{K$_{0.95}$}\\
MAG\_ZEROPOINT    & 24.7576 & mag$_{AB}$  \textsl{K$_{0.95}$}\\
SATUR\_LEVEL      &  100.0  & DN/s \textsl{K$_{2.68}$}\\
ZERO\_POINT       & 23.7848 & mag$_{AB}$   \textsl{K$_{2.68}$}\\
SATUR\_LEVEL      &  100.0  & DN/s \textsl{K$_{\rm spec}$}\\
ZERO\_POINT       & 23.5815 & mag$_{AB}$   \textsl{K$_{\rm spec}$}\\
\enddata
\end{deluxetable*}

\begin{figure*}[htb!]
\gridline{\fig{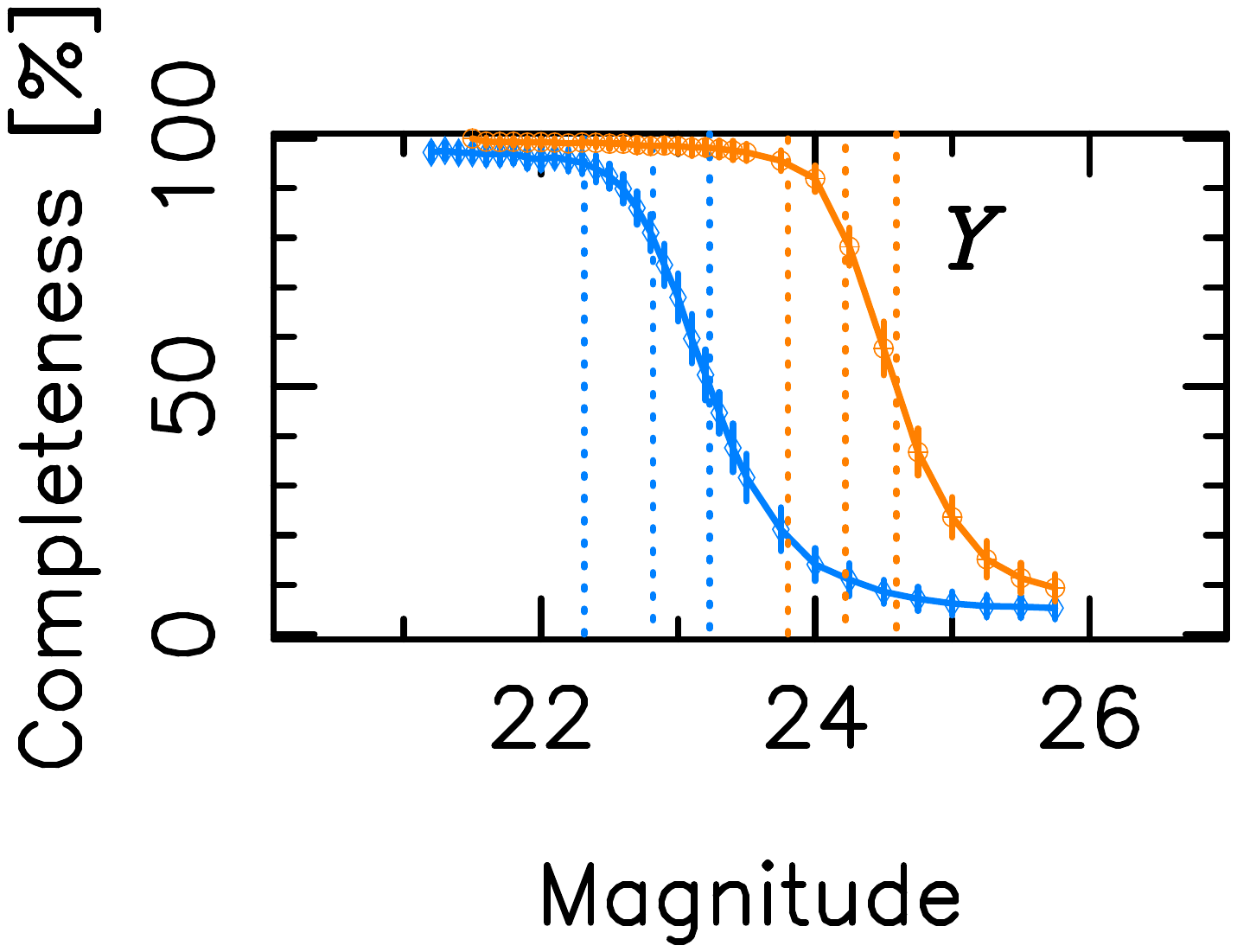}{0.25\textwidth}{(a)}
          \fig{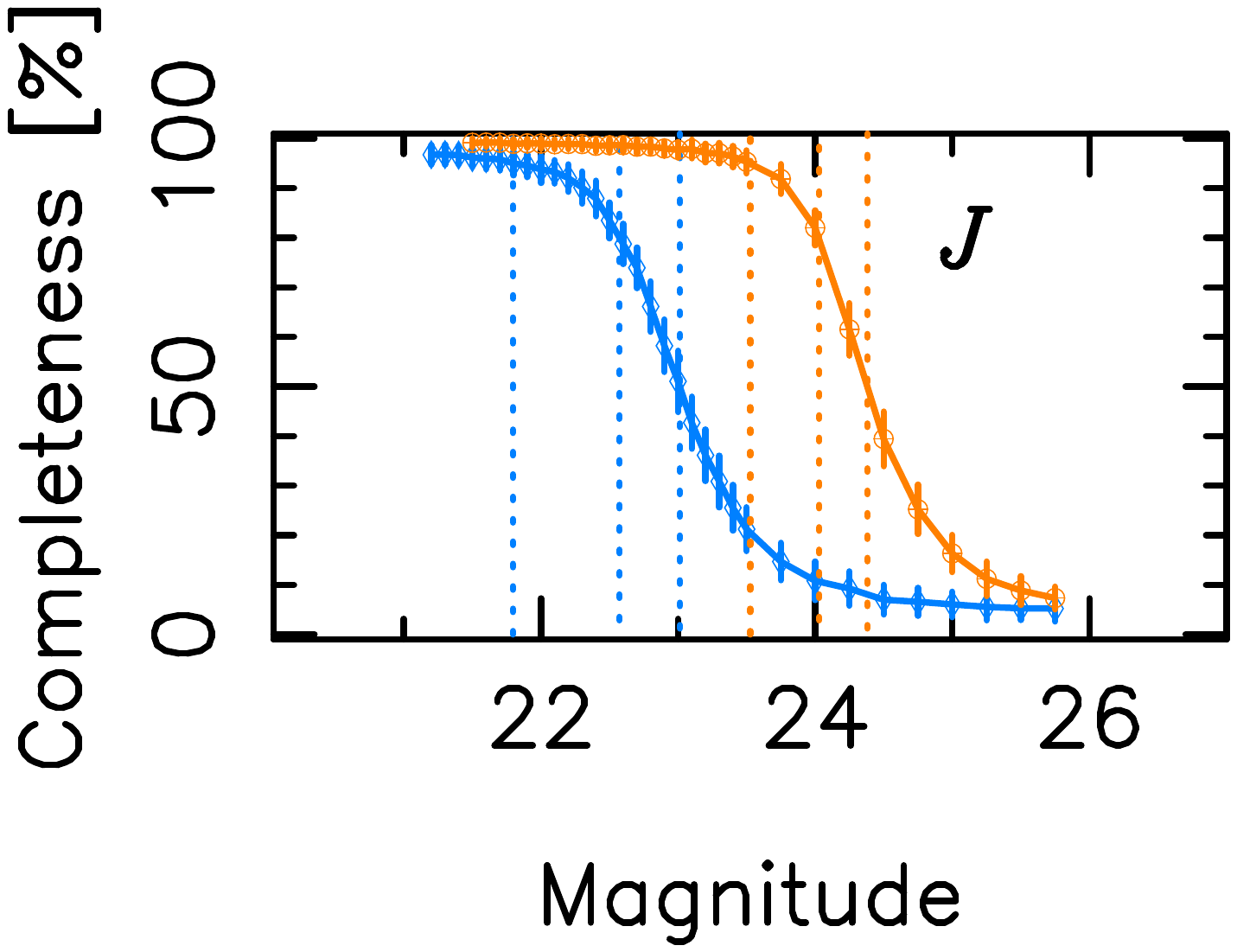}{0.25\textwidth}{(b)}
          \fig{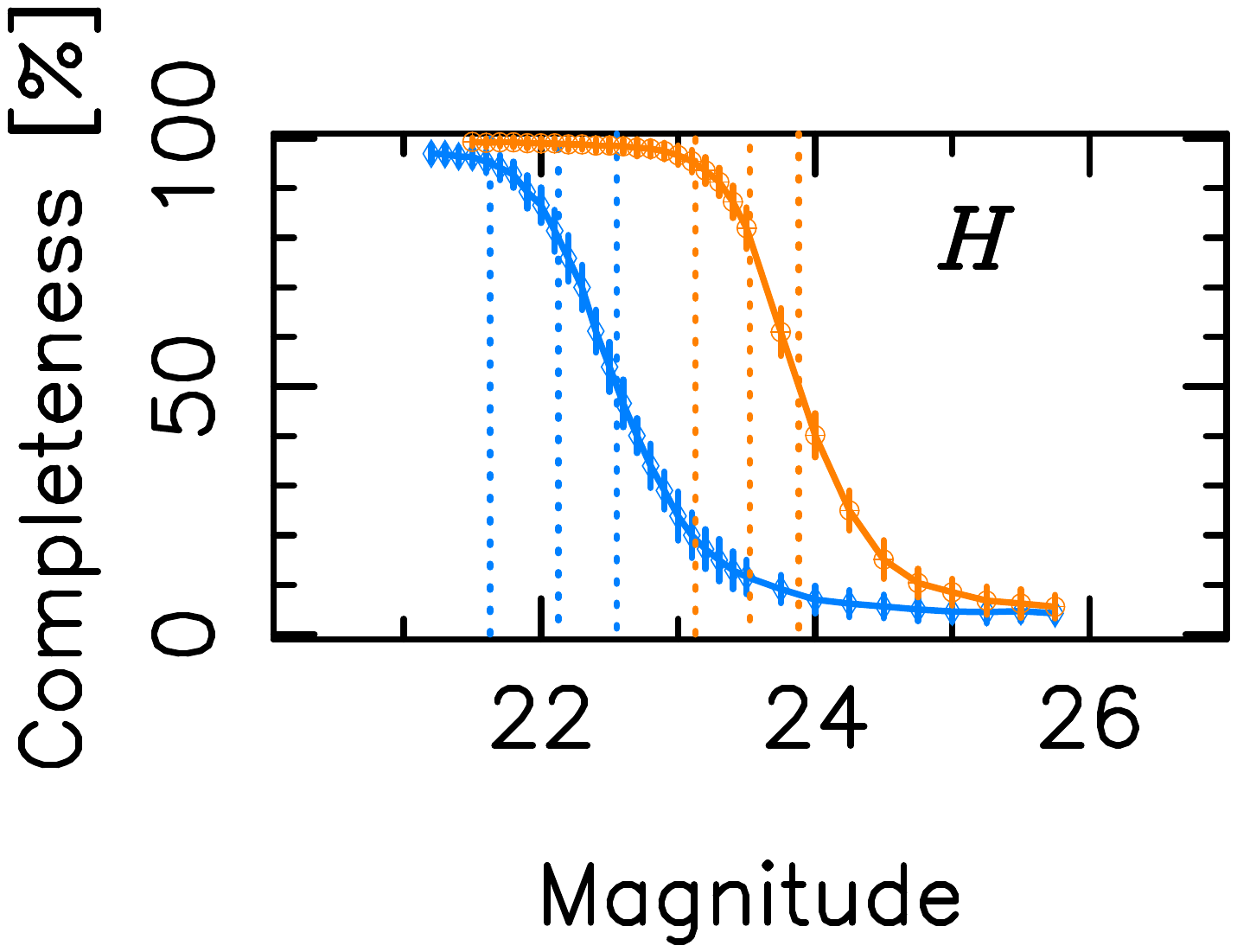}{0.25\textwidth}{(c)}}
\gridline{\fig{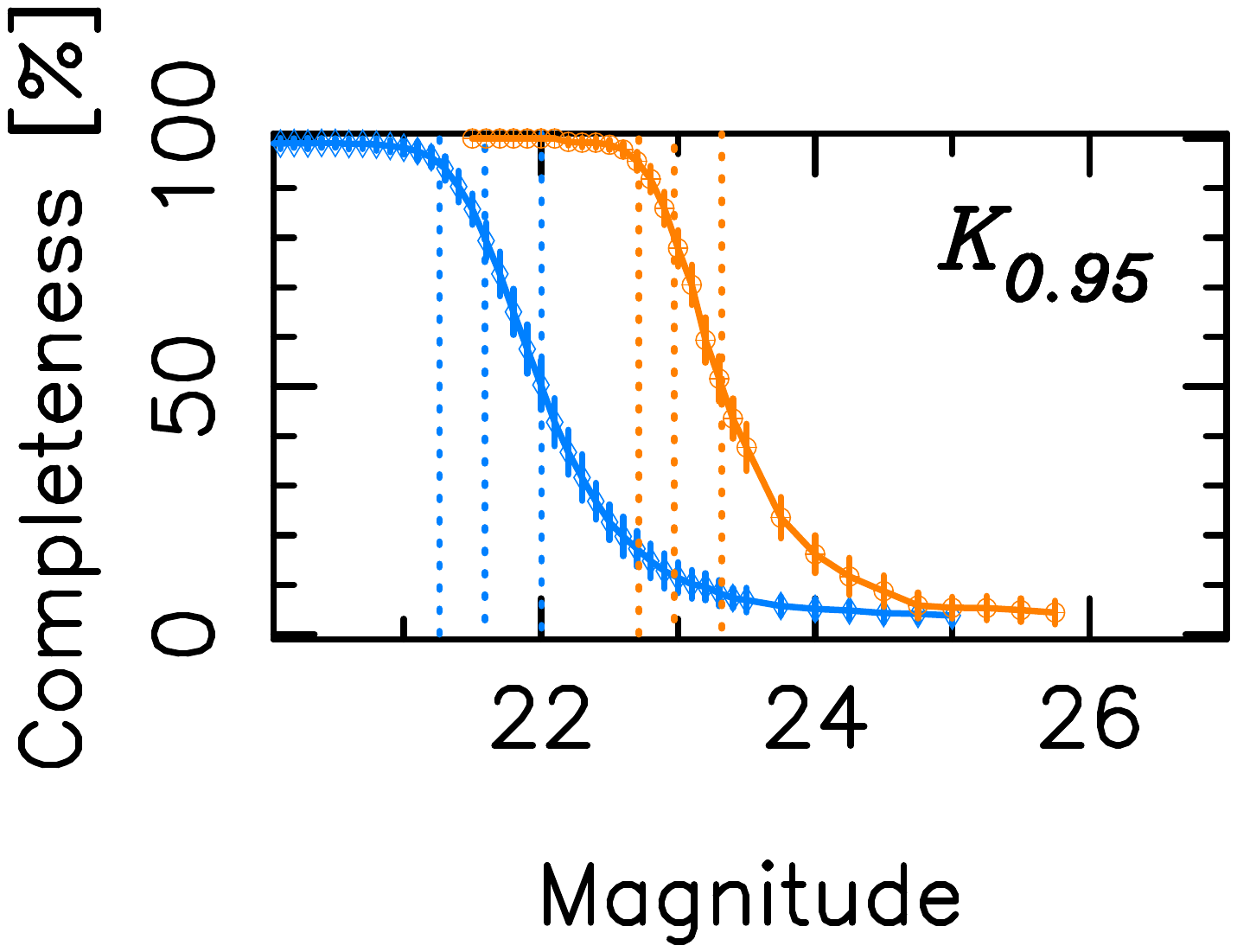}{0.25\textwidth}{(d)}
          \fig{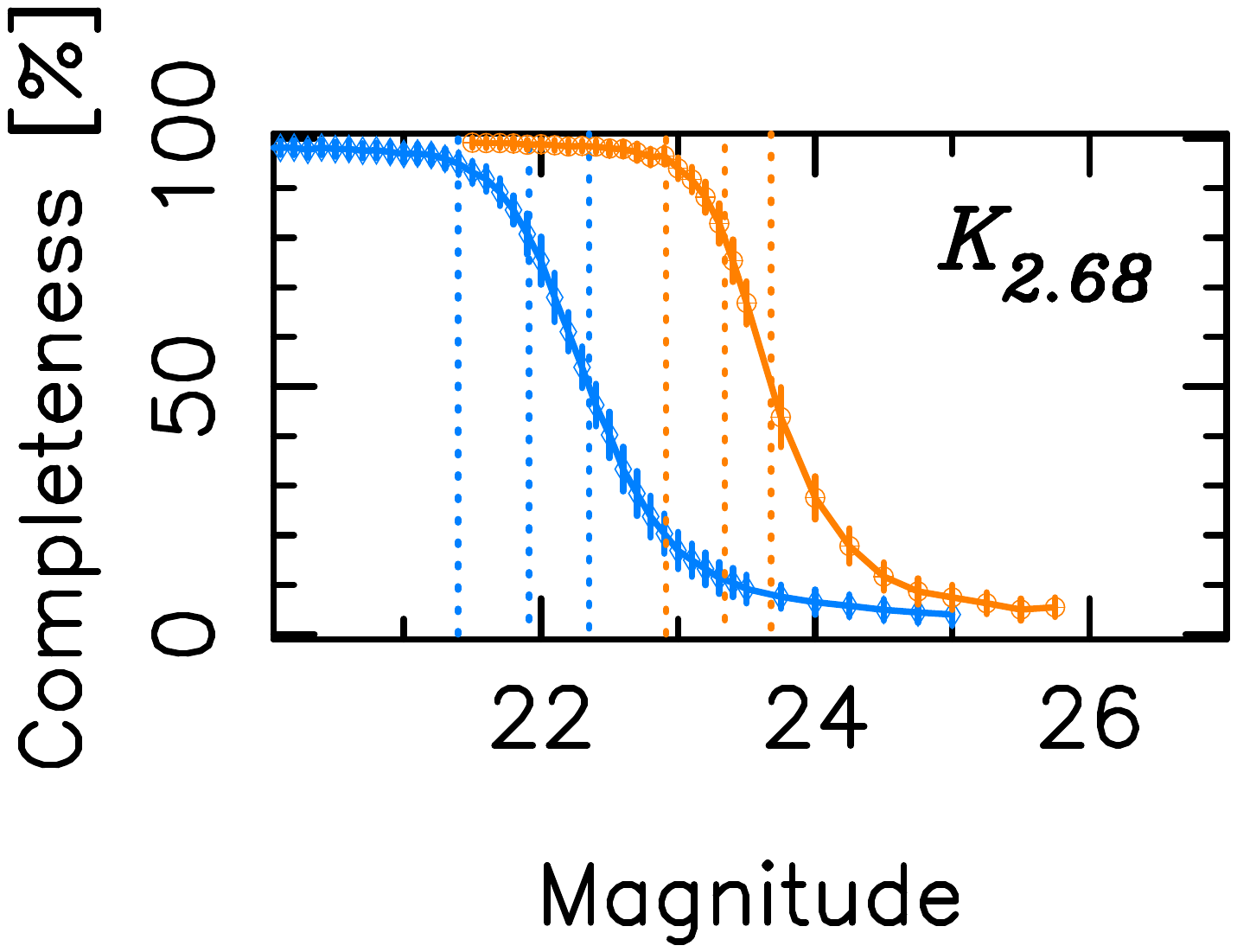}{0.25\textwidth}{(e)}
          \fig{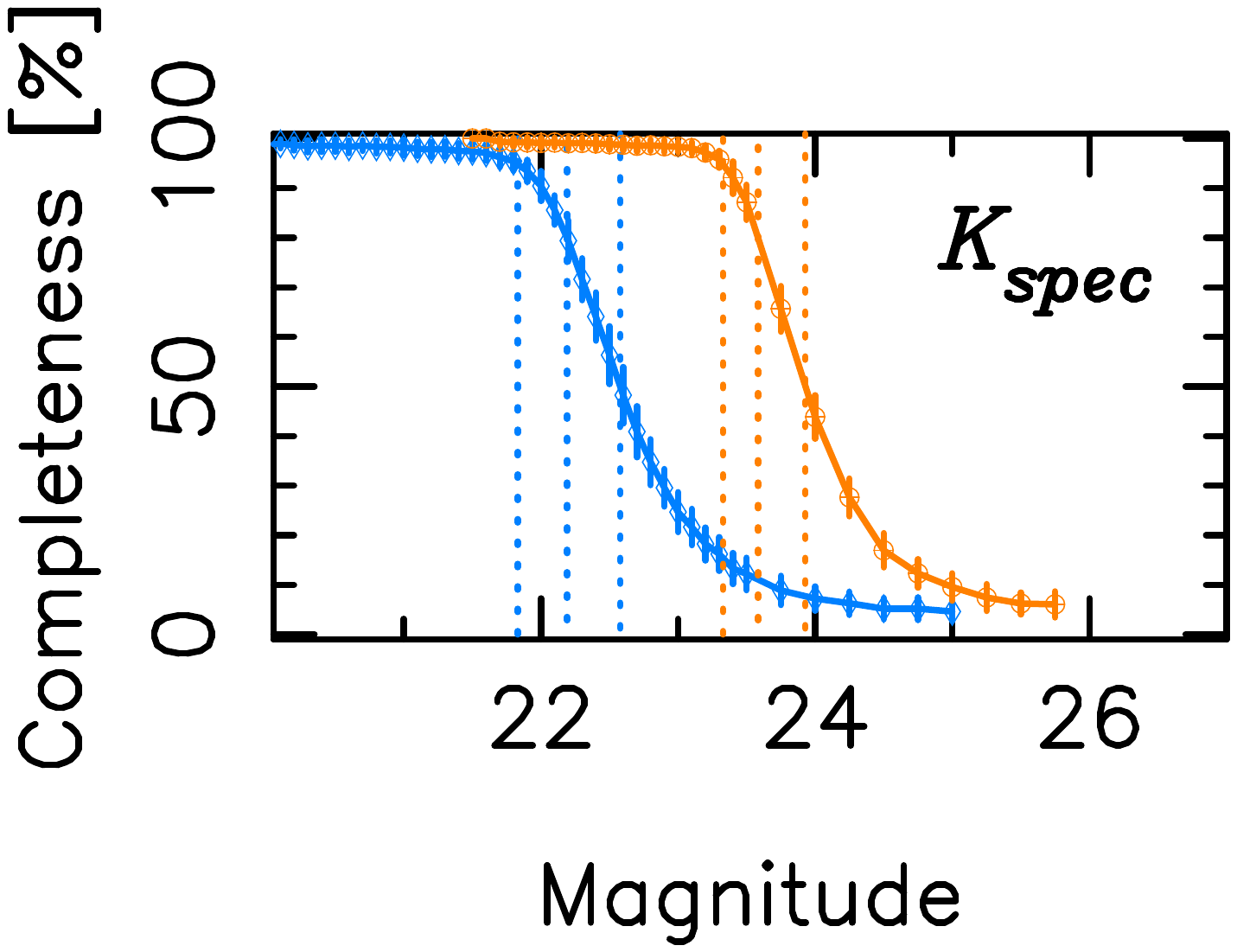}{0.25\textwidth}{(f)}
          \fig{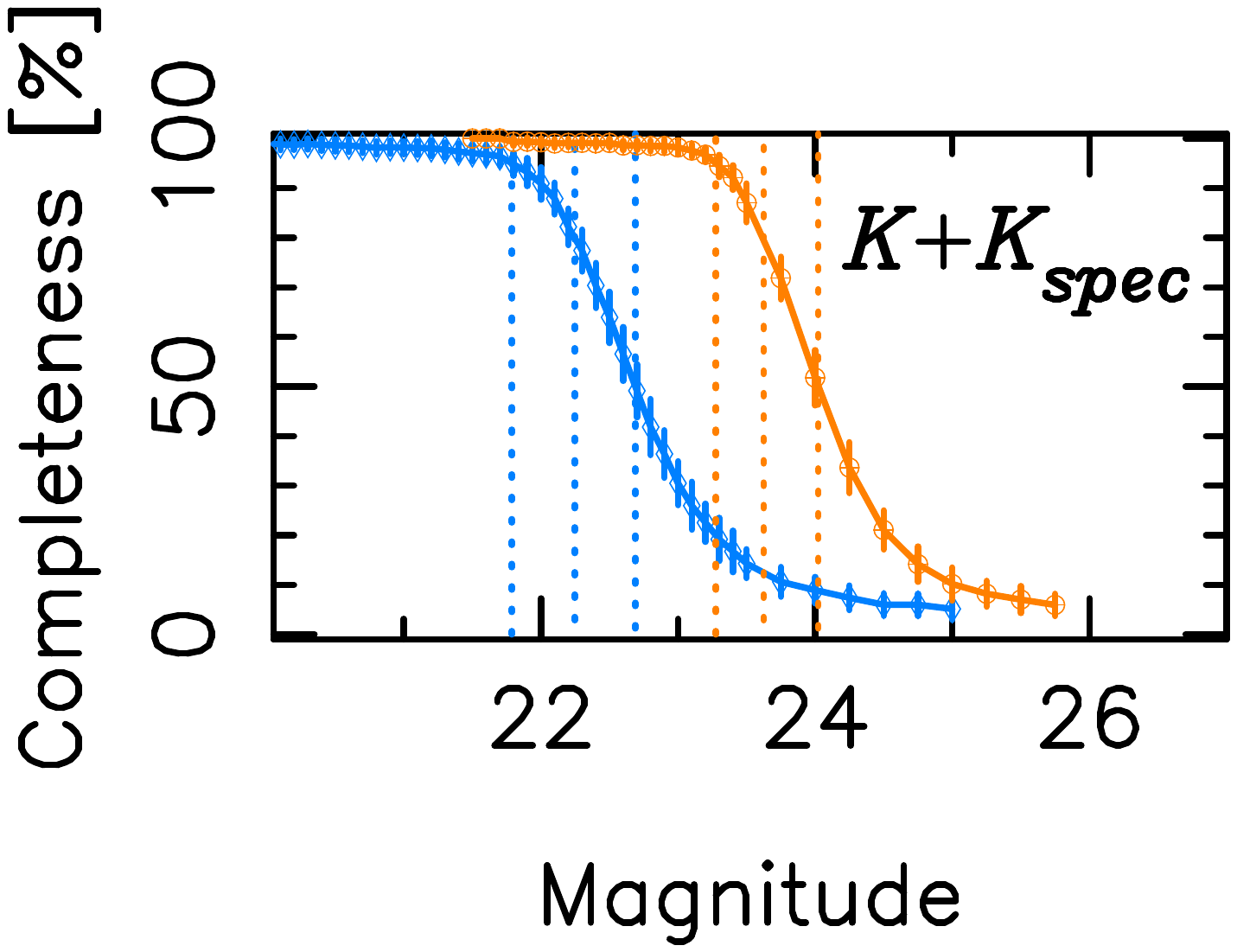}{0.25\textwidth}{(g)}}
\caption{Completeness levels for the individual bands and for the combined \textsl{K+K$_{\rm spec}$} mosaic (\emph{g}) for point sources (orange points and lines) and for $n$\,=\,1 S\'ersic profiles (blue).
The vertical lines indicate (from left to right) the  95\%, 80\%, and 50\% completeness levels.\label{fig:completeness}}
\end{figure*}
\begin{deluxetable*}{lrrrl}[htb!]
\tablecaption{Completeness levels.\label{tab:completeness}}
\tablewidth{0pt}
\tablehead{
\colhead{Filter} & \colhead{m$_{AB}$(95\%)} & \colhead{m$_{AB}$(80\%)} & 
\colhead{m$_{AB}$(50\%)} &\colhead{Source type, minarea\tablenotemark{a}, threshold\tablenotemark{b}}}
\startdata
\textsl{Y} &    23.80 & 24.22 & 24.59 & point source, 10, 1\\
\textsl{Y} &    22.32 & 22.82 & 23.23 & S\'ersic $n$=1,   10, 1\\
\textsl{J} &    23.53 & 24.03 & 24.38 & point source, 10, 1\\
\textsl{J} &    21.80 & 22.57 & 23.01 & S\'ersic $n$=1,   10, 1\\
\textsl{H} &    23.13 & 23.52 & 23.88 & point source, 10, 1\\
\textsl{H} &    21.63 & 22.13 & 22.55 & S\'ersic $n$=1,   10, 1\\
\textsl{K$_{ 0.95}$} &  22.71 & 22.97 & 23.32 & point source, 10, 1\\
\textsl{K$_{0.95}$}&21.26 & 21.59 & 22.00 & S\'ersic $n$=1,   10, 1\\
\textsl{K$_{2.68}$} &22.91 & 23.34 & 23.68 & point source, 10, 1\\
\textsl{K$_{2.68}$}&21.39 & 21.91 & 22.35 & S\'ersic $n$=1,   10, 1\\
\textsl{K$_{\rm spec}$} & 23.33 & 23.58 & 23.93 & point source, 10, 1\\
\textsl{K$_{\rm spec}$}& 21.83 & 22.19 & 22.58 & S\'ersic $n$=1,   10, 1\\
\textsl{K+K$_{\rm spec}$} &23.28 & 23.62 & 24.02 & point source, 10, 1\\
\textsl{K+K$_{\rm spec}$}& 21.79 & 22.25 & 22.69 & S\'ersic $n$=1,   10, 1\\ 
\enddata
\tablenotetext{a}{\texttt{SExtractor} minimum detection area (in pixels$^2$ of 0\farcs168 per pixel) .}
\tablenotetext{b}{\texttt{Sextractor} detection threshold in terms of
standard deviations above the background level.}
\end{deluxetable*}
\section{Results}
\subsection{Catalog completeness}
To create the catalogs for each MMIRS filter, we used exposure
maps to remove sources contained in regions covered by fewer than (5, 4, 7, 9, 9, 9) images for \textsl{Y, J, H, K$_{0.95}$, K$_{2.68}$, K$_{\rm spec}$} because of the low $S/N$ and greater likelihood of including spurious detections. These regions are those outlined in Figure~1.

The completeness of individual bands and for the \textsl{K+K$_{\rm spec}$} mosaic was estimated following, e.g., \citet{Caldwell2006, Finkelstein2015}, by adding simulated sources to cutouts of the science images.
We used sections of 1000$\times$1000 pixels with the exception of the \textsl{K+K$_{\rm spec}$} mosaic, for which a  2000$\times$2000 region was used, centered on the position where all \textsl{K} imaging setups overlap.
Using a grid of 0.1\,mag steps, 100 simulations per magnitude bin were run, where in each simulation 100 sources were added at random positions and catalogs created using the same procedure as for the science catalogs, i.e., using a minimum area of 10 pixels with a detection threshold at 1\,$\sigma$ above the mean background and using the \texttt{SExtractor} \textsf{mag\_auto} estimator for the total magnitude and the detection parameters presented in Table~\ref{tab:se_params}.
Table~\ref{tab:completeness} lists the completeness limits in each band estimated for  point sources modelled by the PSF constructed by \texttt{PSFEX} from the mosaics and for sources with a S\'ersic $n=1$ profile, using \texttt{GalSim} \citep{Rowe2015} models convolved with these PSFs. The S\'ersic models were sampled from a uniform distribution of position angle, half-light radius and axial ratio ranging between [0$^\circ$, 360$^\circ$], [0.8$''$, 3.3$''$] and [0.3, 1) respectively.
Table~\ref{tab:completeness} shows the completeness limits (in AB magnitudes) at the 95\%, 80\%, and 50\% levels for each of the filters (and gain value in the case of \textsl{K}) and for the combined \textsl{K+K$_{\rm spec}$} mosaic.
The completeness curves from which these limits were computed are presented in Figure~\ref{fig:completeness}.
\subsection{Building a combined \textsl{K} sample}\label{sec:kdiff}

\begin{figure}[htb!]
\plotone{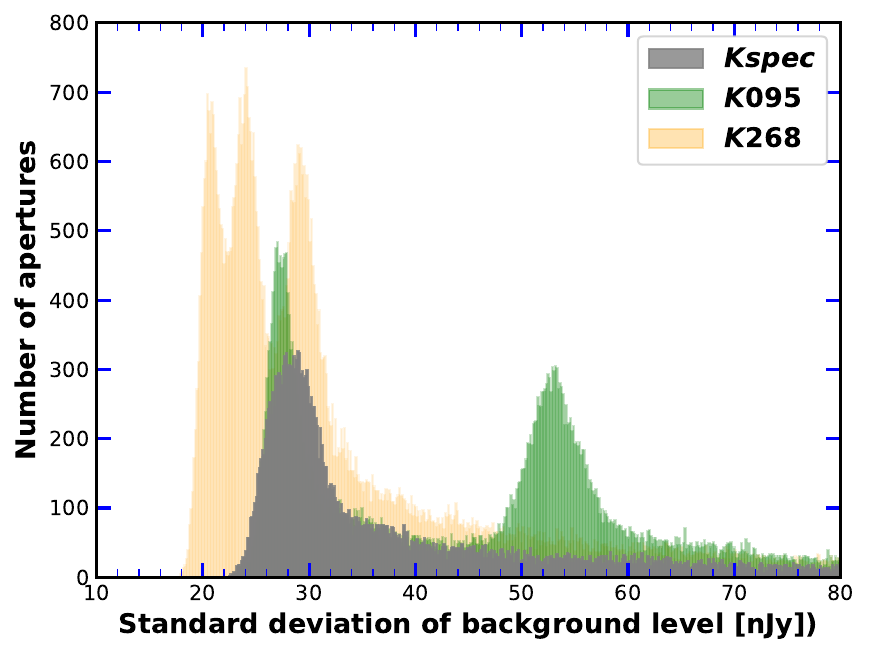}
\caption{Distribution of the dispersion of measured background fluxes using 100,000 random apertures of 12.5 pixels (2\farcs1) calculated for the final science-grade mosaics in \textsl{K$_{2.68}$} (orange), \textsl{K$_{0.95}$} (green) and \textsl{K$_{\rm spec}$} (grey). The number of apertures contributing to the histogram for each filter is proportional to area with exposure times greater than 600s and comprise 78\%, 53\% and 31\% of the random variates for \textsl{K$_{2.68}$}, \textsl{K$_{0.95}$} and \textsl{K$_{\rm spec}$} respectively. The higher noise level for the \textsl{K$_{0.95}$} is immediately apparent; the different peaks are associated to the large contiguous regions of the mosaics with very similar exposure times. Shorter exposures shift the peaks towards higher noise  values as expected; the peaks seen for \textsl{K$_{2.68}$} (from left to right) correspond to total exposures $\sim$ 9100~s, 8600~s and 3800~s, while for \textsl{K$_{0.95}$} the peaks correspond to $\sim$ 4000~s and 1000~s; for \textsl{K$_{\rm spec}$} the counts peak at $\sim$ 6000~s.\label{fig:kfilter_noise}}
\end{figure}

The different filter throughputs and detector gain values have a small, but detectable effect on the data quality of the mosaics. We show in Figure~\ref{fig:kfilter_noise} the distribution of the standard deviation of the background flux measured from 100,000 randomly placed 2\farcs1 apertures in each of the \textsl{K} band mosaics. The \textsl{K$_{2.68}$} measurements (orange distribution) show the smallest dispersion, though there is large amount of overlap between these and the measurements using the wider \textsl{K$_{\rm spec}$} filter (grey) and the \textsl{K$_{0.95}$} (green) setups. Because of this, we created catalogs for the individual instrument settings. 
The different peaks seen for the \textsl{K$_{2.68}$} and \textsl{K$_{0.95}$} setups are due to a large number of pixels that have very similar exposure times and  are concentrated at the center of the mosaics in given quadrant.
%
\begin{deluxetable*}{lrrrrcl}[htb!]
\tablecaption{\textsl{K} magnitude differences\label{tab:kmags}}
\tablewidth{0pt}
\tablehead{
\colhead{Filter and gain setup} & \colhead{mean difference} & \colhead{uncertainty\rlap{\tablenotemark{a}}} &
\colhead{$\sigma$} &
\colhead{Matching sources} & \colhead{Bright (B) or Full (F) sample} &\colhead{Quadrants}
}
\startdata
\textsl{K$_{2.68}$}--\textsl{K$_{0.95}$} & $-$0.031 & 0.007 & 0.282 & 1613~~~~~~~~~ & B &SE, SW \\
\textsl{K$_{2.68}$}--\textsl{K$_{\rm spec}$}      &  0.073 & 0.017 & 0.418 &  621~~~~~~~~~ & B &SE\\
\textsl{K$_{0.95}$}--\textsl{K$_{\rm spec}$}      &  0.122 & 0.010 & 0.340 & 1284~~~~~~~~~ & B &SE\\
\textsl{K$_{2.68}$}--\textsl{K$_{0.95}$} & $-$0.010 & 0.007 & 0.307 & 1921~~~~~~~~~ & F &SE, SW\\
\textsl{K$_{2.68}$}--\textsl{K$_{\rm spec}$}\tablenotemark{b} & 0.130 & 0.018 & 0.513 & 776~~~~~~~~~ &F&SE\\
\textsl{K$_{0.95}$}--\textsl{K$_{\rm spec}$} & 0.212 & 0.011 & 0.442 & 1686~~~~~~~~~ & F & SE\\
\enddata
\tablenotetext{a}{The uncertainty is calculated by
\textbf{resistant\_mean} as $\sqrt{\sigma}/(N-1)$}
\tablenotetext{b}{Both sets use gain of 2.68 e$^-$/DN}
\end{deluxetable*}

\begin{figure*}[htb!]
\plotone{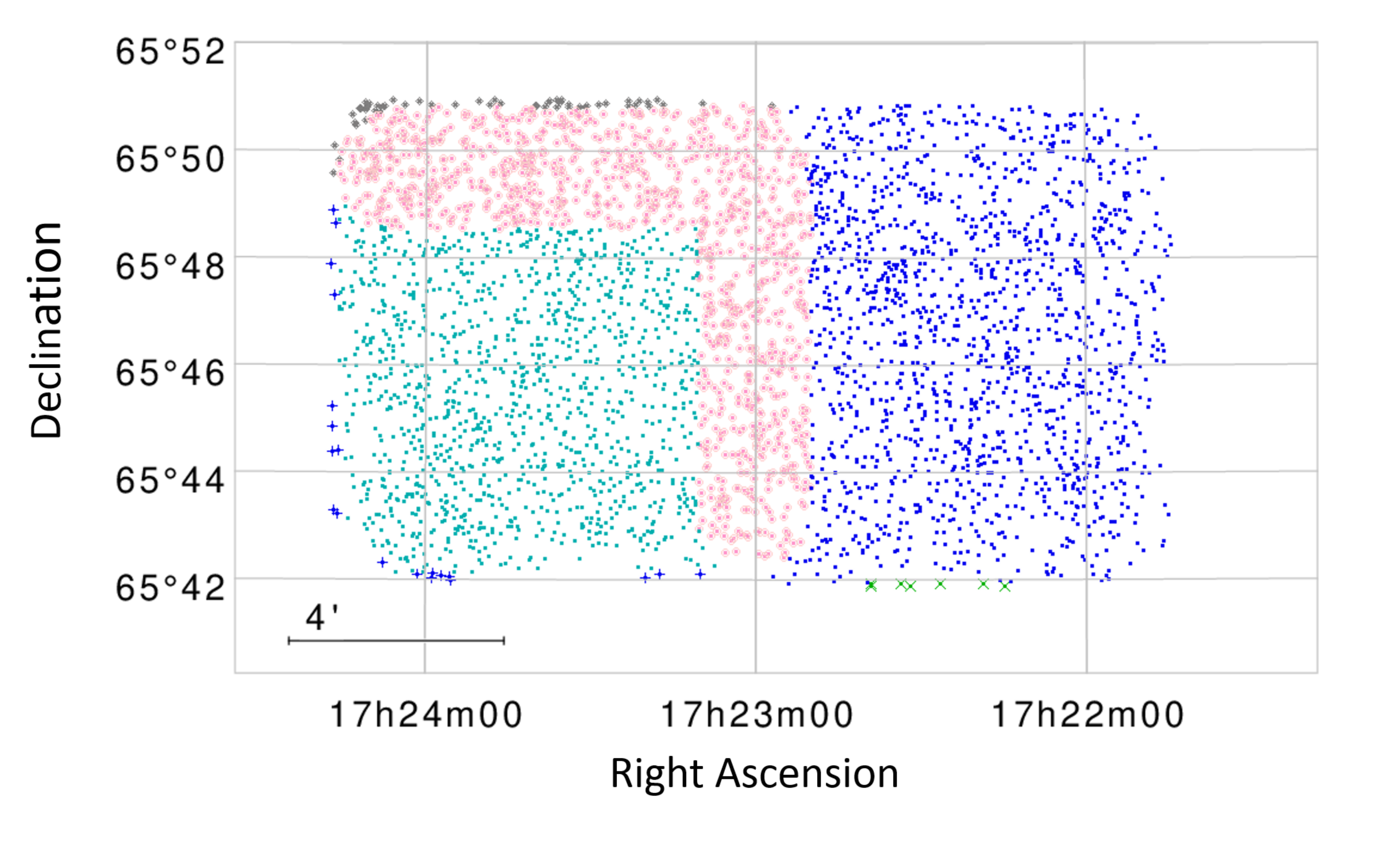}
\caption{Distribution in right ascension and declination of sources in the regions where the three modes of \textsl{K} band imaging overlap. Pink circles show sources in common between all modalities (\textsl{K$_{2.68}$}, \textsl{K$_{0.95}$}, and \textsl{K$_{\rm spec}$}). Blue points represent sources with \textsl{K$_{2.68}$} and \textsl{K$_{0.95}$} imaging; green  points sources with \textsl{K$_{0.95}$} and
  \textsl{K$_{\rm spec}$}. Sources detected in \textsl{K$_{\rm spec}$} and \textsl{K$_{0.95}$} only are represented by black ``+'s'' and green ``x's'' respectively. Grey diamonds represent sources with \textsl{K$_{\rm spec}$} and \textsl{K$_{2.68}$} measurements.\label{fig:kmag}}
\end{figure*}


The regions where each of these settings overlap 
were used to estimate average offsets between the different imaging samples (Table~\ref{tab:kmags} and Figure~\ref{fig:kmag}).
In this calculation, we used the \texttt{IDL} task \textbf{resistant\_mean} with a cutoff at 3.5\,$\sigma$ and restricted the sample to sources with \texttt{Sextractor} flags = 0. We considered two cases: in the first one, we used sources with magnitudes in the range 18 $\leq$ \textsf{mag\_auto} $\leq$ 23 mag to exclude bright saturated sources and faint sources where the photometry becomes uncertain; the second case repeated the same calculation but  without magnitude cuts. 
From Table~\ref{tab:kmags}, using the bright sample, we found an offset of $-$0.031 $\pm$ 0.007 mag between \textsl{K$_{2.68}$} and \textsl{K$_{0.95}$}. The
offset between  \textsl{K$_{0.95}$}$-$\textsl{K$_{\rm spec}$} is 0.122 $\pm$ 0.010 mag. 
Combining the \textsl{K$_{2.68}$}$-$\textsl{K$_{0.95}$} and \textsl{K$_{0.95}$}$-$\textsl{K$_{\rm spec}$} differences implies that \textsl{K$_{2.68}$}$-$\textsl{K$_{\rm spec}$} $\sim$ 0.091 mag.
The direct measurement of \textsl{K$_{2.68}$}$-$\textsl{K$_{\rm spec}$} = 0.073 mag suggests that constructing a single magnitude sample can have systematic uncertainties of the order of $\sim$0.02 magnitudes.

To create a uniform sample, we used \textsl{K$_{2.68}$} as reference magnitudes, subtracted 0.031 mag from \textsl{K$_{0.98}$} and added 0.073 mag to the \textsl{K$_{\rm{spec}}$} measurements. The final magnitude for each source  is an inverse variance-weighted average using the corrected fluxes in each modality with weights derived from the estimated flux uncertainty measured by \texttt{SExtractor}.

\begin{figure*}[htb!]
\gridline{
{\fig{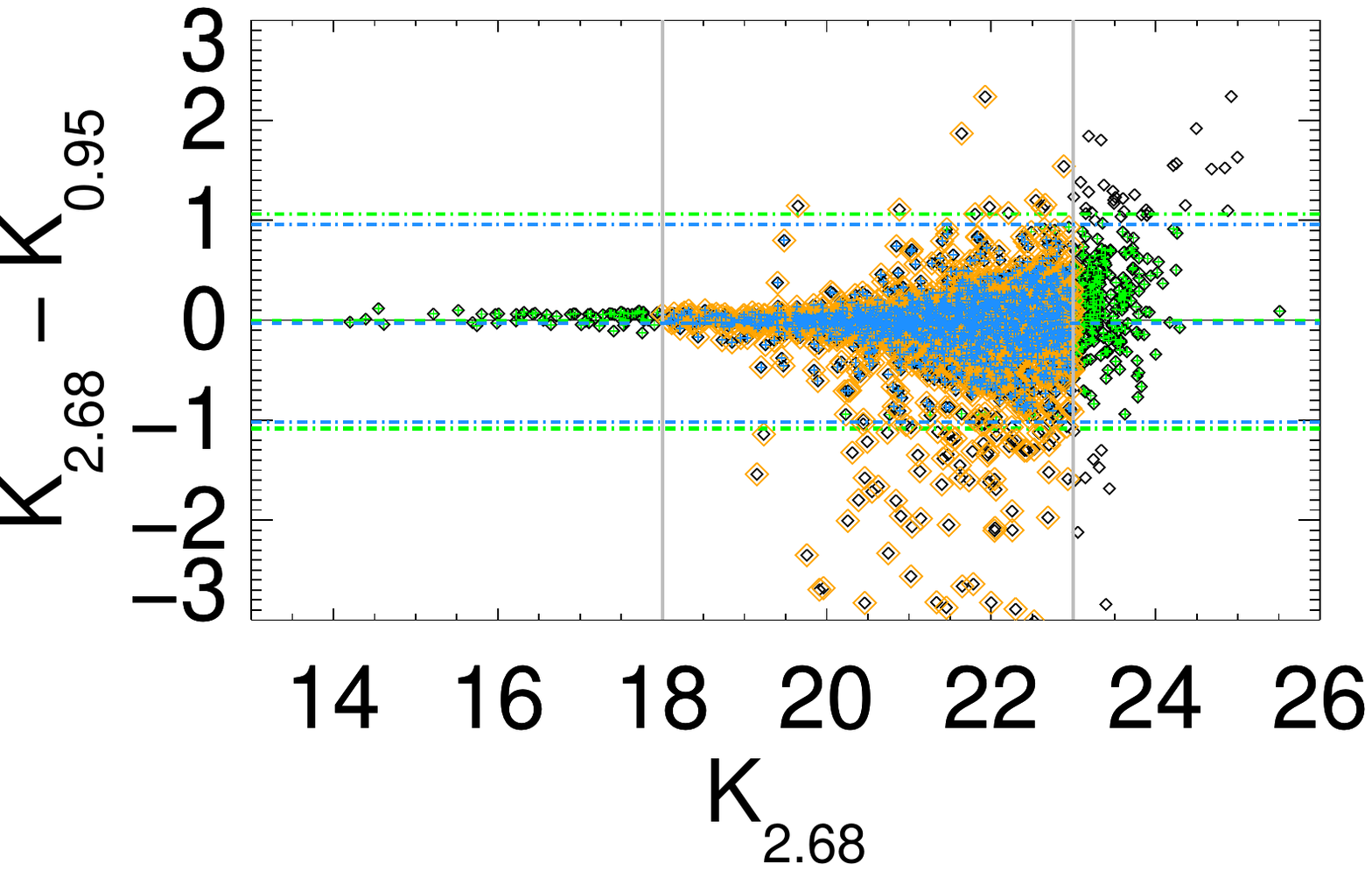}{0.31\textwidth}{(a)}}
{\fig{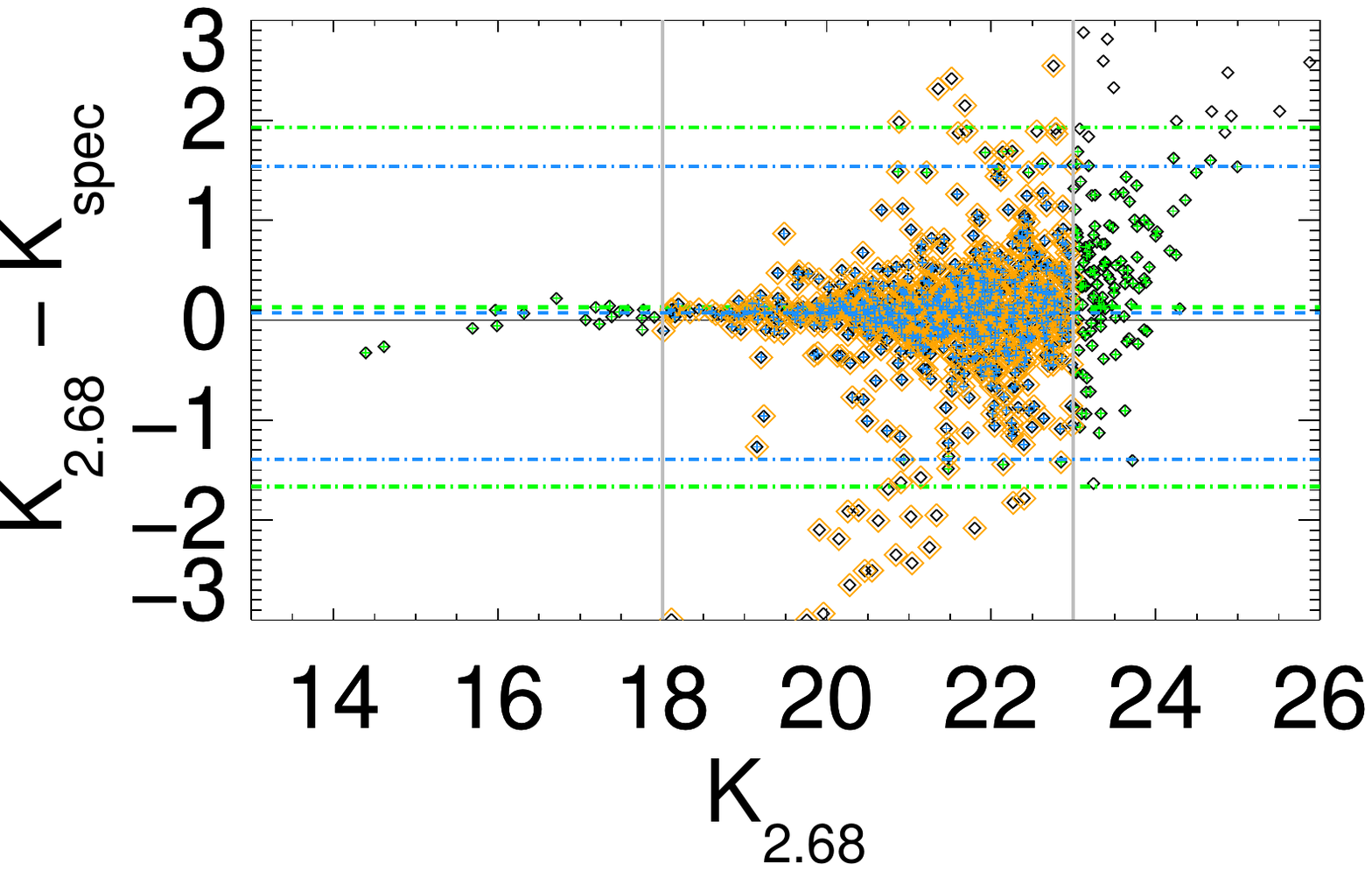}{0.31\textwidth}{(b)}}
{\fig{K095_Kspec_comp.pdf}{0.31\textwidth}{(c)}}
}
\caption{Comparisons between photometric measurements of sources for the different \textsl{K} band filters and gain values. The full sample of galaxies, with no magnitude cut, is plotted as black diamonds; green dots represent sources that remain after a 3.5$\sigma$ cut. The orange diamonds represent the sub-sample of galaxies after the magnitude cuts and black plusses the same sample after a 3.5$\sigma$ cut. The dashed lines represent the mean calculated using the \textbf{resistant\_mean} procedure in \texttt{IDL} and the horizontal dot-dashed lines the $\pm$1\,$\sigma$ range for the full sample (green) and the magnitude-limited one (blue). The vertical dashed lines represent the bright and faint magnitude limits used to calculate the average offsets.\label{fig:k_comp}}
\end{figure*}

\subsection{Archival Subaru Hyper-Suprime-Cam data}\label{subsec:hsc}

To expand the wavelength coverage and provide visible counterparts to the NIR sources, we used archival Subaru Hyper-Suprime-Cam (HSC) data that were available in early 2020 from the SMOKA\footnote{\url{https://smoka.nao.ac.jp/}} archive maintained by the National Astronomical Observatory of Japan. These images were obtained as part of HEROES (Songaila et al. \citeyear{Songaila2018}, Taylor et al. \citeyear{Taylor2023}) and comprise data of a single field -- \textsl{NEP-wide-A05}. The observation log is shown in Table~\ref{tab:hsc_log}, which lists
the filter name, total exposure time, 
the limiting magnitude estimated from the peak of the magnitude distribution in 0.5 mag bins, filter pivot wavelength and bandwidth \citep[e.g.,][]{2018ApJS..236...47W}, 
the flux corresponding to a count of 1 photon per second, and dates of observation.
\begin{deluxetable*}{lccrrcl}[htb!]
\centering 
\tablecaption{\textsl{HSC} Observation log.\label{tab:hsc_log}}
\tablewidth{0pt}
\tablehead{
\colhead{filter} & 
\colhead{exposure} & 
\colhead{m$_{\rm lim}$}  &
\colhead{$\lambda_{pivot}$\tablenotemark{a}} &
\colhead{d$\lambda$\tablenotemark{a}} &
\colhead{F$_{\nu0}$\tablenotemark{a}}&
\colhead{UT observation date(s)}\\
\colhead{} & 
\colhead{sec} & 
\colhead{AB magnitude} &
\colhead{$\mu$m}&
\colhead{$\mu$m}&
\colhead{erg cm$^{-2}$ Hz$^{-1}$ s$^{-1}$}&
{}
}
\startdata
\textsl{g}      &  525 & 25.1 & 0.4780 & 0.1208 & 4.06004$\times$10$^{-20}$ & 2017-06-29\\
\textsl{i2}     & 1200 & 24.6 & 0.7735 & 0.1495 & 2.58121$\times$10$^{-20}$ & 2017-06-23, 2017-08-23\\
\textsl{z}      & 1200 & 24.2 & 0.8910 & 0.0778 & 2.31604$\times$10$^{-20}$ & 2017-06-28\\
\textsl{NB0816} & 2000 & 24.7 & 0.8177 & 0.0113 & 2.40826$\times$10$^{-20}$ & 2017-06-29, 2017-08-26\\
\textsl{NB0921} & 1200 & 24.1 & 0.9214 & 0.0135 & 2.25802$\times$10$^{-20}$ & 2017-06-22\\
\enddata
\tablenotetext{a}{Values from \citet{2018ApJS..236...47W}, {\url{http://mips.as.arizona.edu/~cnaw/sun.html}}}
\end{deluxetable*}

The data were reduced using the HSC pipeline \citep{Bosch2018}, and for the \TDF\ analyses, a region of $\sim$30.80$'$\,$\times$\,33.59$'$ centered at (17:22:48.1298, +65:50:14.853, J2000) was excised, which contains the footprints of JWST (PI: R.~Windhorst \& H.~Hammel) and HST (PI: R.~Jansen \& N.~Grogin) imaging and most of the other ancillary data, e.g., JCMT/SCUBA2 and VLA, \citep{Hyun2023}; Chandra (PI: W.~Maksym); NuSTAR (PI: F.~Civano; \citealt{Zhao2021}) and XMM-Newton (PI: F.~Civano).  The source catalog was created using \texttt{SExtractor} \citep{BertinArnouts} in dual-imaging mode with a detection image stacking the
\textsl{i2}, \textsl{z}, \textsl{NB0816} and \textsl{NB0921} images
following \citet{Szalay1999}. The detection was run in two steps - ``cold'' and ``hot'' where some of the detection parameters are changed to minimize the fragmentation of large and bright galaxies; these parameters are presented in Table~\ref{tab:hsc_params} and those that change according to the detection mode are noted.
The final PEARLS HSC catalog is a single list concatenating the ``cold'' and ``hot'' 
output from the five individual bands and contains data for 56786  sources. 
The astrometric and photometric calibration used the same process described by \citet{Bosch2018} and \citet{Aihara2022} with the Pan-STARRS DR2 as reference and is tied to the  GAIA DR1 astrometric reference frame \citep{GAIADR1}. 
The HSC images are calibrated into maggies{\footnote{\url{https://www.sdss4.org/dr12/algorithms/magnitudes/}}}, which are in the AB system \citep{Oke_Gunn_1983}, 
and the flux value corresponding to  magnitude 0 is specified by the FLUXMAG0 image header keyword\footnote{\url{https://hsc.mtk.nao.ac.jp/pipedoc/pipedoc_8_e/tips_e/read_catalog.html}} .
The conversion from maggies into Jy or erg~cm$^{-2}$~s$^{-1}$~Hz$^{-1}$ requires multiplying the image values by 10$^{8.9/2.5}$ or 10$^{-48.6/2.5}$ respectively.
Because of the flux units adopted by the HSC pipeline, the \texttt{SExtractor} magnitudes are calculated directly in the AB system. However, the expression used by \texttt{SExtractor} to calculate the magnitude uncertainties: 
\begin{widetext}
\begin{equation}
  {\rm magerr} = \frac{2.5}{\ln(10)} \frac {\sqrt{{\rm area} \times {\rm variance(background)} +
    {\rm flux}/{\rm gain}}} {\rm {flux}}
\end{equation}
\end{widetext}
assumes flux measurements in DN/s (data numbers per second), and will produce incorrect uncertainties
(by several orders of magnitude) because the statistical uncertainty in photon counts
has to be separated from the conversion of photon counts into flux units. 
The magnitude uncertainties we report are obtained converting the AB magnitudes
into photon numbers per cm$^2$ per second using  
\begin{equation}
{\rm photon\_numbers} = \frac{f_{\nu0} \times d\lambda} {h \times \lambda}
\end{equation}
where $f_{\nu}0$ is the flux measured in erg cm$^{-2}$ Hz$^{-1}$ s$^{-1}$ corresponding to one event per second in a given filter; $\lambda$ and d$\lambda$ are the filter's characteristic wavelength and bandwidth and $h$ is
Planck's constant. We make the assumption that the instrument gain is already contained in the calibration and use the pivot wavelength (e.g., \citet{BessellMurphy2012}, eq. A13) as the characteristic wavelength.  Assuming
Poisson errors both for background and source photons the expression
above becomes:
\begin{widetext}
\begin{equation}
  {\rm magerr} = \frac{2.5}{\ln(10)} \frac{\sqrt{[{\rm area} \times {\rm variance(background)} +
    {\rm flux}] \times {\rm photon\_numbers} \times {\rm mirror~area}}} {{\rm flux} \times {\rm photon\_numbers} \times {\rm mirror~area}},
\end{equation}
\end{widetext}
where area is the total number of pixels covered by the source, and  flux and the background variance are values measured by \texttt{SExtractor}.
Using the filter parameters in Table~\ref{tab:hsc_log} and an effective
diameter of 820 cm for the Subaru telescope\footnote{
\url{https://subarutelescope.org/Observing/Telescope/Parameters/}},
this expression produces estimated uncertainties of the same order of
magnitude as those available in the SDSS database for objects in common, and are slightly more conservative than the uncertainty estimates in the full HEROES catalog \citep{Taylor2023}, discussed in the next paragraph. Both flux and flux uncertainties are converted into nJy by multiplying each value by 10$^{31.4/2.5}$.

\begin{deluxetable*}{lll}[htb!]
\tablecaption{HSC \textsl{SExtractor} detection parameters.\label{tab:hsc_params}}
\tablewidth{0pt}
\tablehead{
\colhead{Parameter} & \colhead{value} & \colhead{note}}
\startdata
DETECT\_THRESH    & 5.0    & $\sigma$ above mean background cold mode\\
ANALYSIS\_THRESH  & 5.0    & $\sigma$ above mean background cold mode\\
DETECT\_MINAREA   & 10     & pixels \\
DEBLEND\_NTHRESH  & 32     & {} \\
DEBLEND\_MINCONT  & 0.001  & cold mode\\
{}                &  {}    &  {}\\
DETECT\_THRESH    & 1.10   & $\sigma$ above mean background hot mode\\
ANALYSIS\_THRESH  & 1.10   & $\sigma$ above mean background hot mode\\
DETECT\_MINAREA   & 10     & pixels \\
DEBLEND\_NTHRESH  & 32     & {} \\
DEBLEND\_MINCONT  & 0.0001  & hot mode \\
{}                &  {}      & {}\\
FILTER\_NAME      & gauss\_1.5\_3$\times$3.conv\\
PHOT\_FLUXFRAC    & 0.5    & {} \\
PHOT\_AUTOPARAMS  & 2.5,3.5 & Kron factor, minimum radius \\
PHOT\_AUTOAPERS   & 0.0,0.0 & {}\\
CLEAN\_PARAM      & 0.3    & {} \\
WEIGHT\_TYPE      & BACKGROUND & {} \\
RESCALE\_WEIGHTS  & N       & {} \\
WEIGHT\_THRESH    & 1.e-08  & {} \\
SEEING            & 0\farcs7 & {}\\
SATUR\_LEVEL      & 8.0e-09 & maggies\\
\enddata
\end{deluxetable*}

A catalog for the full HEROES survey was published by \citet{Taylor2023}, who also used the pipeline described by \citet{Bosch2018} 
to produce a band-merged aperture matched-catalog for the full HEROES region. The HEROES catalog divides the survey into several patches, four of which ({\textbf{17,9}}, {\textbf{17,10}}, {\textbf{18,9}},{ \textbf{18,10}}) cover the \TDF. 

The catalog of unique sources (\textbf{HEROES\_Full\_Catalog.fits}) was used to recover some of the photometric measurements only available in the patch-level catalogs, e.g., aperture magnitudes. We restricted this catalog to the \TDF\ region -- 260.0583 (17:20:13.992) $\leq$ R.A. $\leq$ 261.3350  (17:25:20.400) and 65.550 (65:33:00) $\leq$ Dec. $\leq$ 66.125 (67:07:30) -- and contains 150,216 sources after setting the flags 
\begin{equation}
\texttt{is\_primary} = 1,
\end{equation}
and for each filter
\begin{widetext}
\begin{equation}
\textsl{\{g,i2,z,NB816,NB921\}}\texttt{\_base\_PixelFlags\_flag\_edge} = false, 
\end{equation}
and
\begin{equation}
\textsl{\{g,i2,z,NB816,NB921\}}\texttt{\_base\_PixelFlags\_bad} = false,
\end{equation}
\end{widetext}
following \citet{Taylor2023}.
Matching the latter with the PEARLS catalog of HSC sources 
produces a list of 55,796 objects in common.
To assess the quality of the PEARLS photometry, we selected stellar sources using the distribution of HEROES \texttt{MAG\_PSF} magnitudes versus the \texttt{MAG\_PSF}~-~\texttt{MAG\_CMODEL} magnitude difference following \citet{Bosch2018}. In the comparison that follows, we restricted sources to having
\begin{equation}
        \|\texttt{MAG\_PSF - MAG\_CMODEL}\| \leq 0.01
\end{equation}
for all wide band filters and only considering sources with 17.5 $\leq$ \textsl{i2} $\leq$ 23.5.

This excised  catalog was used to verify the \texttt{SExtractor} photometry  after accounting for some of the procedure differences, e.g., the HSC pipeline uses pixel radii to define  the aperture photometry  (\texttt{SExtractor} uses diameters) and the HSC pipeline adds an aperture correction when producing the final magnitudes for stars and galaxies, but is not used in the aperture magnitudes \citep{Bosch2018}. The photometry comparison used the aperture magnitude measured at a radius of 12 pixels, which corresponds to a 4.03$"$  aperture diameter as this showed the best compromise between the amount of light being measured and the dispersion between both sets of measurements. 
The results from this comparison are shown in Table~\ref{tab:hsc_heroes}, where the mean value and estimated uncertainty of the differences in magnitude are calculated using the \texttt{IDL} \textbf{resistant\_mean} procedure. The last column shows the multiplicative factor that corrects the fluxes measured by \texttt{SExtractor} to be consistent with the HEROES values. 

\begin{deluxetable*}{lrCrc}[htb]
\tablecaption{Magnitude differences between HEROES and this work.\label{tab:hsc_heroes}}
\tablewidth{0pt}
\tablehead{
\colhead{Filter} & \colhead{$N_{\rm stars}$} &\colhead{mean difference} & \colhead{uncertainty} & \colhead{Flux correction}\\
}
\startdata
\textsl{g}     & 2219 & -0.022 &  0.052 &   1.02043\\
\textsl{i2}    & 2219 & +0.087 &  0.025 &   0.92318\\
\textsl{z}     & 2219 & -0.012 &  0.025 &   1.01125\\
\textsl{NB816} & 2219 & -0.009 &  0.033 &   1.00789\\
\textsl{NB921} & 2219 & -0.009 &  0.030 &   1.00790\\
\enddata
\end{deluxetable*}
\subsection{Merged Visible-NIR catalog}
To create a merged catalog joining the PEARLS  HSC and MMIRS photometry,
sources were matched using a search radius of 0\farcs8 between PEARLS HSC and the NIR sources and between the different NIR catalogs.
While the position uncertainties of the MMIRS and PEARLS HSC are smaller than this -- 0\farcs070 and 0\farcs030 respectively -- the larger search radius allows matching the larger and brighter objects. Cases where there were multiple matches were visually inspected and resolved in the cases of obvious mis-matches.
The final list of unique sources includes those with matches as well as sources with detections in single NIR bands. The total number of  sources in the final merged catalog is 57501 (including 56752 PEARLS HSC sources and respectively (\textsl{Y}, \textsl{J}, \textsl{H}, \textsl{K}) = (8128, 10558, 7612, 6361) of which 354 were found to be spurious. 
The number of 
sources with a detection in a single NIR band is (50, 146, 151, 73) for \textsl{Y, J, H, K} respectively.
As the number of single detections is small, these were visually inspected to remove spurious sources due to hot pixels, residual cosmic rays, background fluctuations, and stellar diffraction spikes. The final tally is (1, 32, 82, 35) sources in each of the NIR filters, but we note that several sources have PEARLS HSC counterparts which are below the catalog detection threshold. The larger number of single detections in \textsl{H} is due to the use of the combined \textsl{H} and \textsl{K} images in the source detection and the better image quality of the observations (e.g., Figure~\ref{fig:seeing}). 
Once the spurious detections are removed, the remaining number of sources is (\textsl{Y}, \textsl{J}, \textsl{H}, \textsl{K}) = (7990, 10355, 7535, 6316), and the total number of bona fide sources is 57467, of which 715 are NIR-detected without counterparts in the PEARLS HSC catalog.

\begin{figure}[htb!]
\centering
\epsscale{0.70}
\plotone{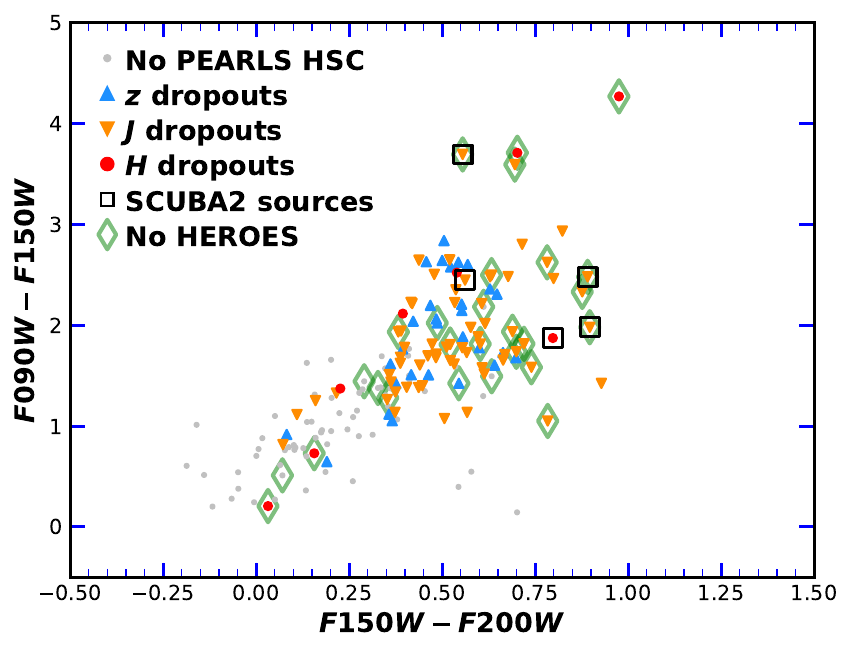}
\caption{Color-color plot of sources matching the PEARLS HSC-MMIRS with a preliminary JWST-TDF catalog measured from the first two \TDF\ spokes. The magnitudes of the latter were measured using \texttt{SExtractor} with a fixed aperture determined from the \textsl{F444W} filter. The 174 plotted NIR sources have no HSC counterparts and the key identifies the different characteristics, i.e., sources that have at least one NIR measurement (grey dots; sources with measurements in all NIR filters 
 (blue upward pointing triangles); sources with measurements  in \textsl{H} and \textsl{K} (orange downward pointing triangles) and only in \textsl{K} (red circles). The 5 galaxies which match SCUBA2 sources are noted by open black squares and the 28 MMIRS-detected galaxies with no HEROES counterparts are represented by green open diamonds.\label{fig:jwst_drops}}
\end{figure}

The number of sources without PEARLS HSC detections\footnote{i.e., a counterpart may exist in the PEARLS HSC imaging, but is below the catalog detection threshold.} 
but present in all NIR (\textsl{Y, J, H}, and \textsl{K}) catalogs (``\textsl{z}-dropouts'') is 118; those with \textsl{H} and \textsl{K} (``\textsl{J}-dropouts'') is 323. The latter comprise candidate objects close to bright sources which were not deblended correctly in the PEARLS HSC catalog (20 sources, $\sim$ 6\% of the unmatched subsample), sources with PEARLS HSC counterparts below the detection level (96 sources, $\sim$29.5\%), and legitimate higher-redshift sources (210 galaxies, $\sim$64.5\%). Given the relatively bright limits of these samples, these ``dropouts'' are most likely caused by the Balmer Break shifting through the HSC and MMIRS filters.
To better understand the properties of these sources without PEARLS
HSC matches, we matched the PEARLS HSC-MMIRS catalog with a
preliminary catalog of the \TDF\ comprising the first two spokes,
which contains 24,119 sources. Using a 0.50$"$ matching radius we
identify 1858 counterparts, 1419 of which have MMIRS measurements in
at least one filter. Figure~\ref{fig:jwst_drops} shows the 174 sources
with no PEARLS HSC counterparts but detected in at least one MMIRS
filter. Of these, 8 have only \textsl{K} measurements (red circles),
66 have \textsl{H} and \textsl{K} (orange triangles) and the remaining
29 are detected in all NIR filters (blue triangles). The grey circles represent the remaining  69 sources that have no HSC detection and have one or two NIR detections but not necessarily in contiguous filters; in one case this is a source contaminated by a diffraction spike.
The visual inspection of these objects suggests the majority as higher-redshift ($z\gtrsim$~1.5) or dusty galaxies -- five of these galaxies are also SCUBA2 sources discovered by \citet{Hyun2023}. One source is a quiescent galaxy at $z~\sim$~2 (N. Adams, private communication). 

We used the HEROES \citep{Taylor2023} subcatalog discussed in \S4.3 to have an independent assessment of the 715 galaxies without PEARLS HSC matches. HEROES includes additional photometry specifically obtained in the \TDF\ with significantly deeper imaging in the \textsl{z} band and additional data in the \textsl{HSC-r2} and \textsl{HSC-y} bands. We find 562 sources with HEROES counterparts within 0.5 arcseconds while 153 have no HEROES match. Of these, 12 are detected in all MMIRS bands, 93 in \textsl{H, K} and 20 in \textsl{K} only; 5 are detected in \textsl{J} where there are no MMIRS observations in the other filters. There are 17 objects detected in \textsl{Y} and \textsl{J} only and most of these are due to imperfect matches, because of merged images or sources close to bright stars. We also matched this catalog to the JWST catalog of spokes 1 and 2 and we find 28 objects in common, which are identified in Figure~\ref{fig:jwst_drops} as green diamonds; four of these are detected in \textsl{K} only and the majority (16) are detected in both \textsl{H} and \textsl{K}.

The matched PEARLS HSC-MMIRS-JWST catalog was also used to verify if any sources show evidence of variability, comparing measurements in the closest pair of filters (e.g., \textsl{F090W}$-$\textsl{z}, \textsl{F115W}$-$\textsl{J},\textsl{F150}$-$\textsl{H}, and
\textsl{F200W}$-$\textsl{K}). By considering only the unsaturated bright sources, there is no evidence of variability. The sources presenting large magnitude differences are all due to contamination either by diffraction spikes or very close bright objects.

\subsection{Star/Galaxy separation and Number Counts}
\begin{figure*}[htb!]
\gridline{
{\fig{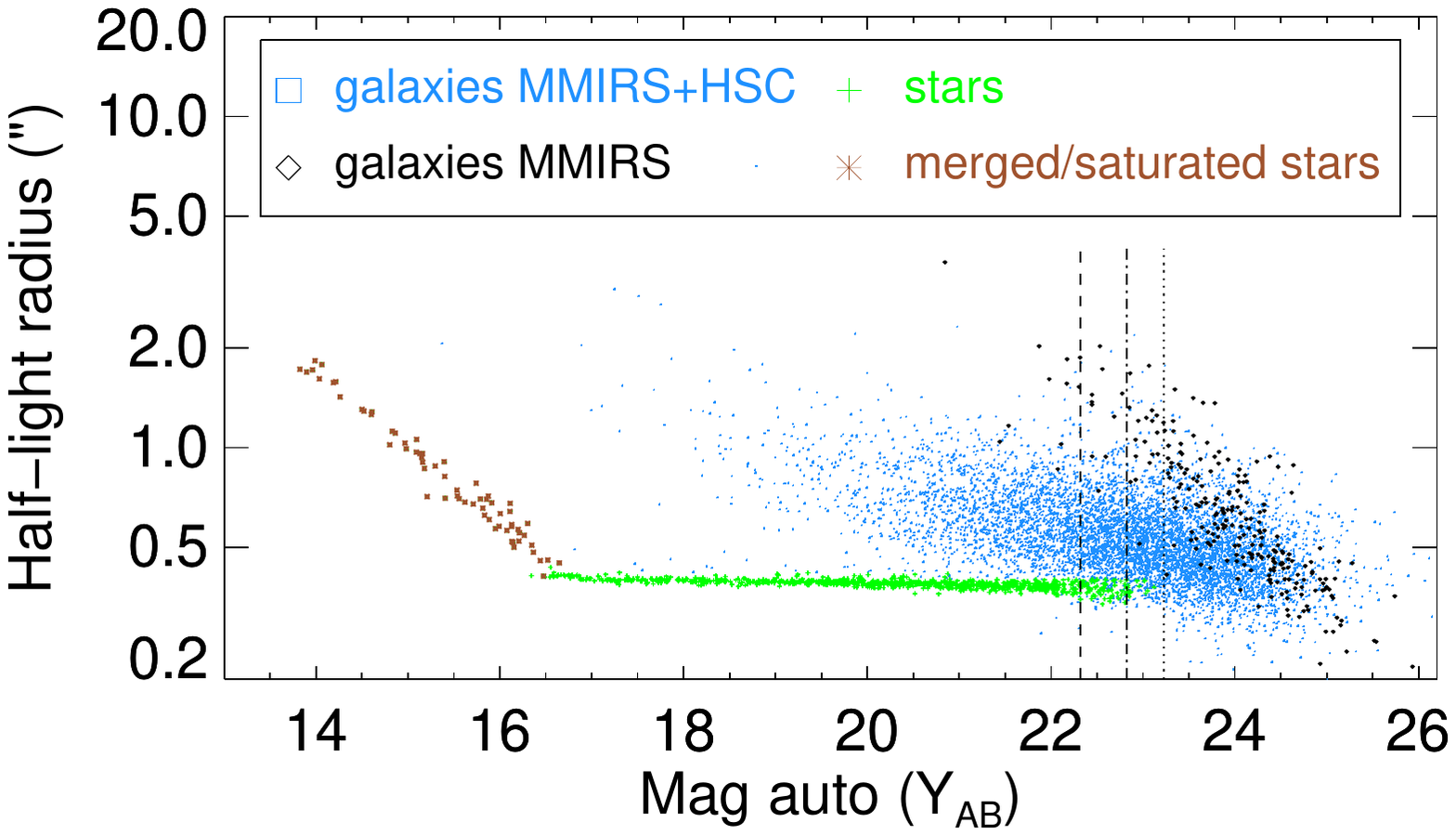}{0.53\textwidth}{(a)}}
{\fig{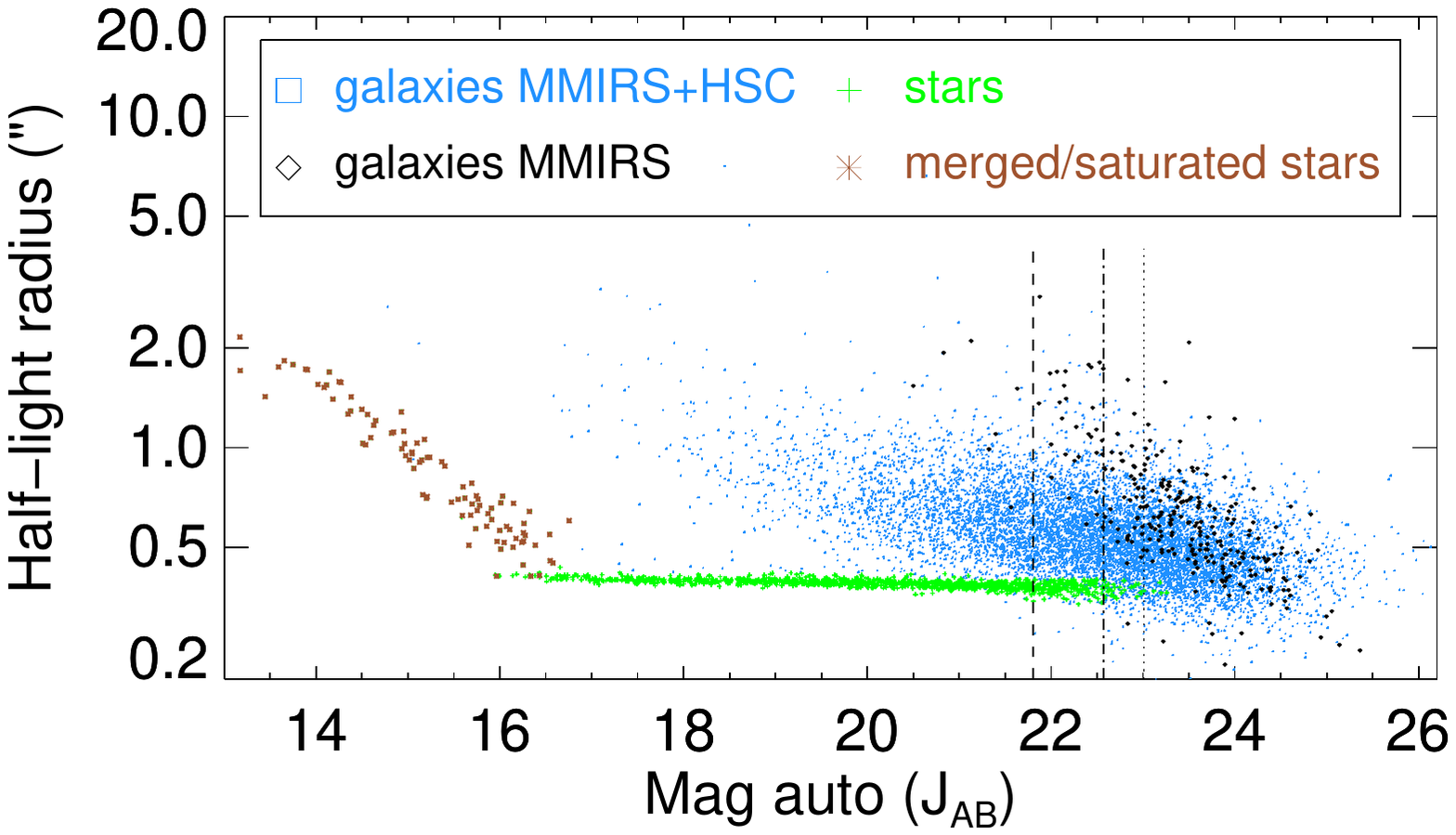}{0.53\textwidth}{(b)}}}

\gridline{
{\fig{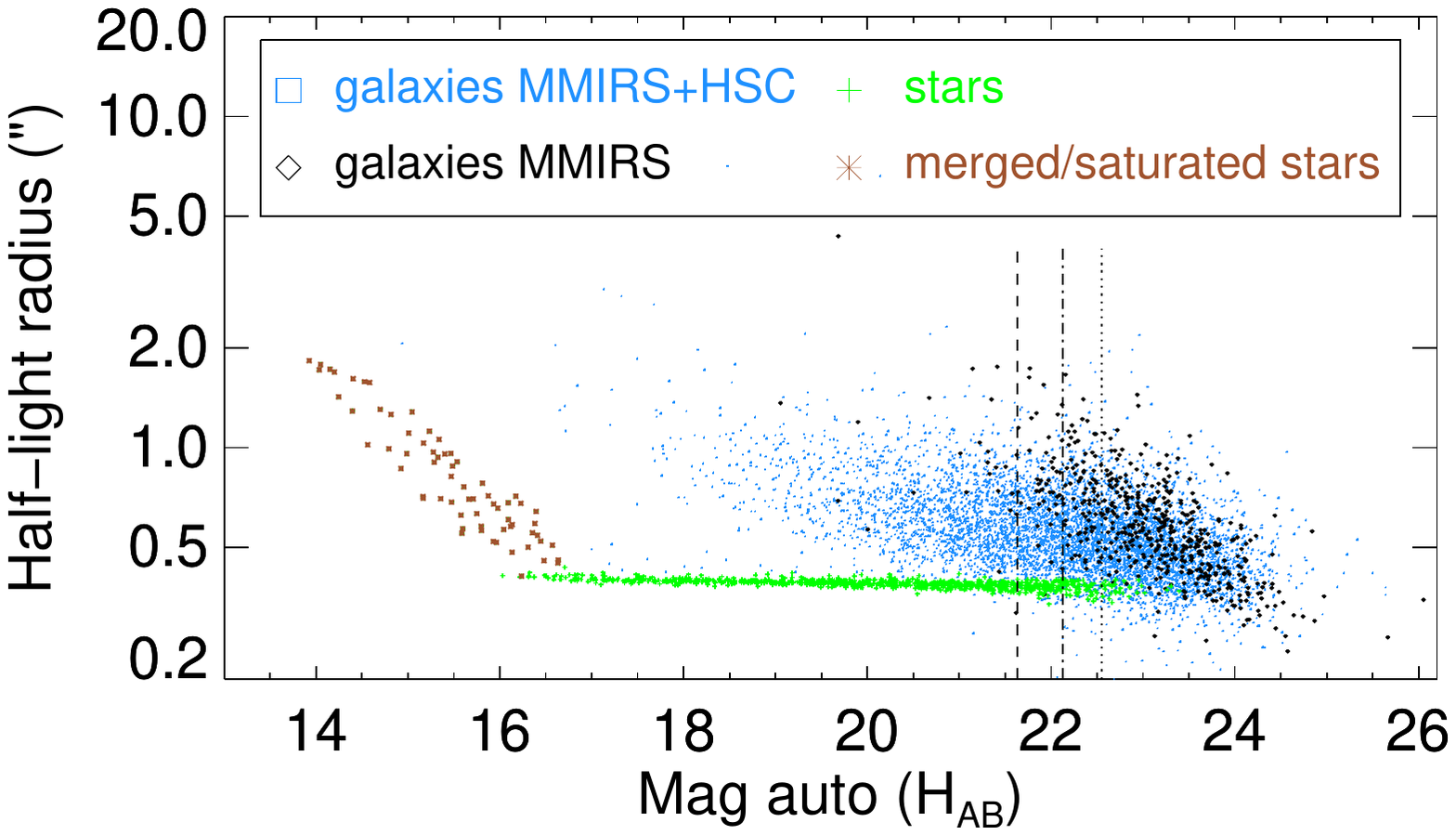}{0.53\textwidth}{(c)}}
{\fig{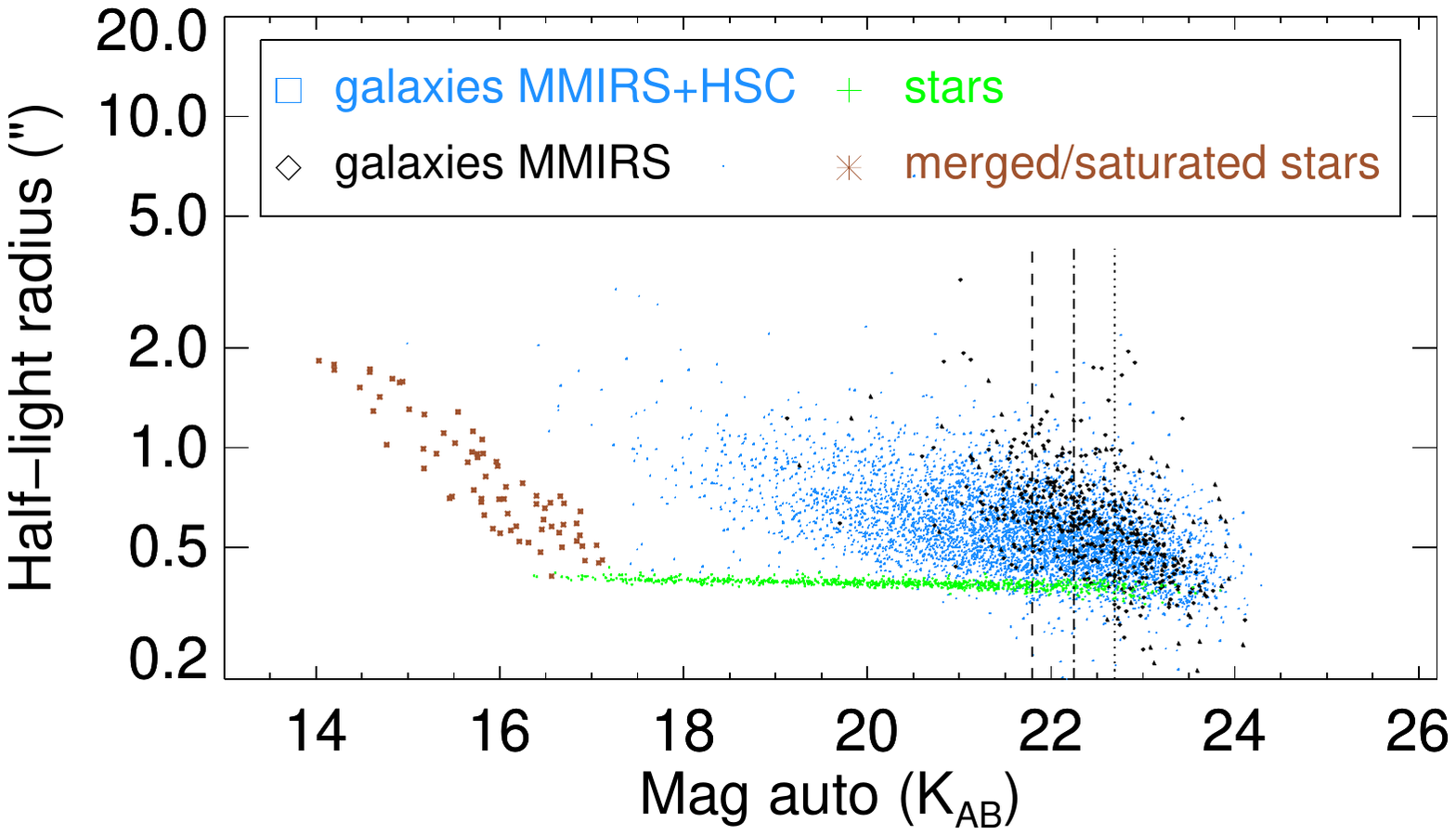}{0.53\textwidth}{(d)}}}
\caption{Star--galaxy separation using the total magnitude (\textsf{mag\_auto}) versus half-light radius measured from the HSC \textsl{z}-band photometry. Green plusses represent sources classified as stars; small blue squares represent galaxies; brown asterisks represent saturated stars or merged images where one or more sources is a star. For sources without an HSC match within 0\farcs80, the NIR half-light radius is used, and these are represented by black diamonds. The vertical lines represent the completeness for extended sources for the 95\% level (dashed line), 80\% (dot-dashed line), and 50\% level (dotted line).\label{fig:star_gal}}
\end{figure*}

The PEARLS HSC+MMIRS merged catalog was used to classify sources detected in the NIR bands as stars or galaxies (or more correctly, compact or extended sources). For this we used the total magnitude versus the half-light radius, e.g., \citet{kron1980}, measured from the HSC \textsl{z} band and the total magnitude (\textsf{mag\_auto})  in each NIR band. We used the radii measured for the PEARLS HSC because of the greater depth and overall better image seeing. In the case of  MMIRS sources without PEARLS HSC half-light radius measurements, we used the corresponding NIR-band half-light radius values. The results from this classification are shown in Figure~\ref{fig:star_gal}, 
where sources of different types are distinguished by symbol shapes and colors as noted in the key of each panel. We also show the 95\%, 80\% and 50\% completeness levels for extended sources as  dashed, dot-dashed and dotted lines. The plots show that below the 80\% completeness level the star/galaxy classification starts breaking down, though this changes with the filter being used.
%
\begin{figure*}[htb!]
\gridline{
{\fig{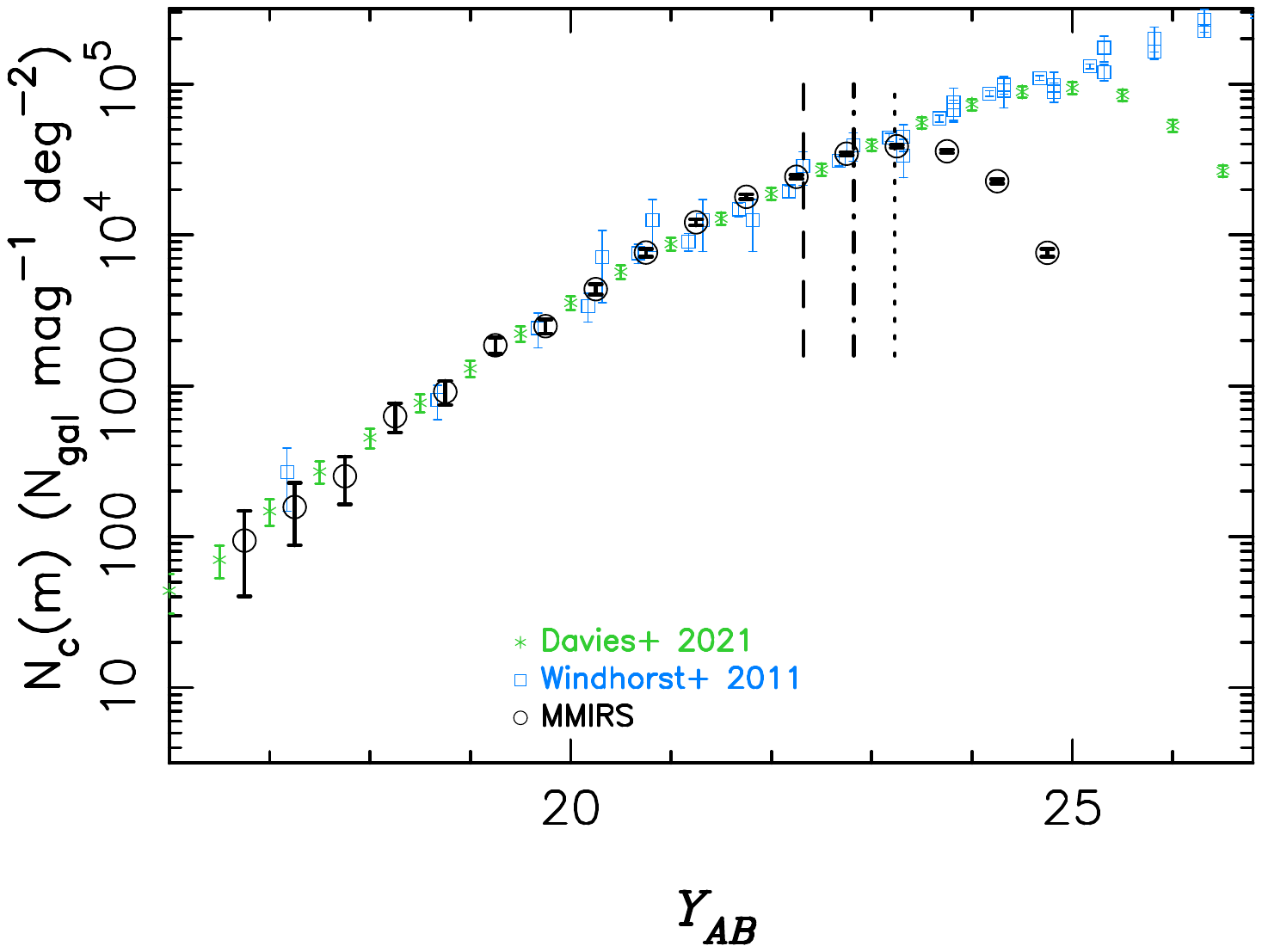}{0.45\textwidth}{(a)}}
{\fig{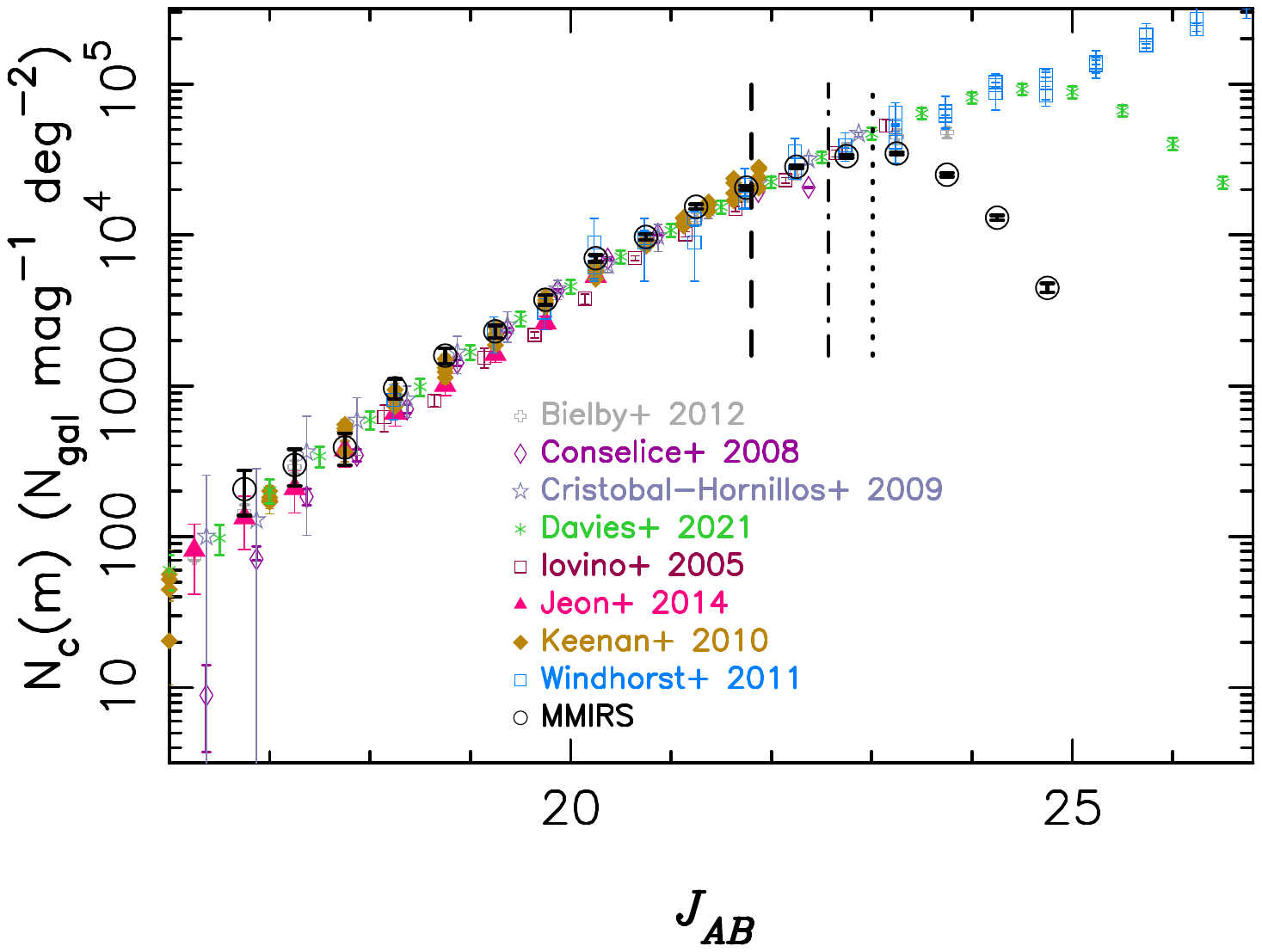}{0.45\textwidth}{(b)}}
}
\gridline{
{\fig{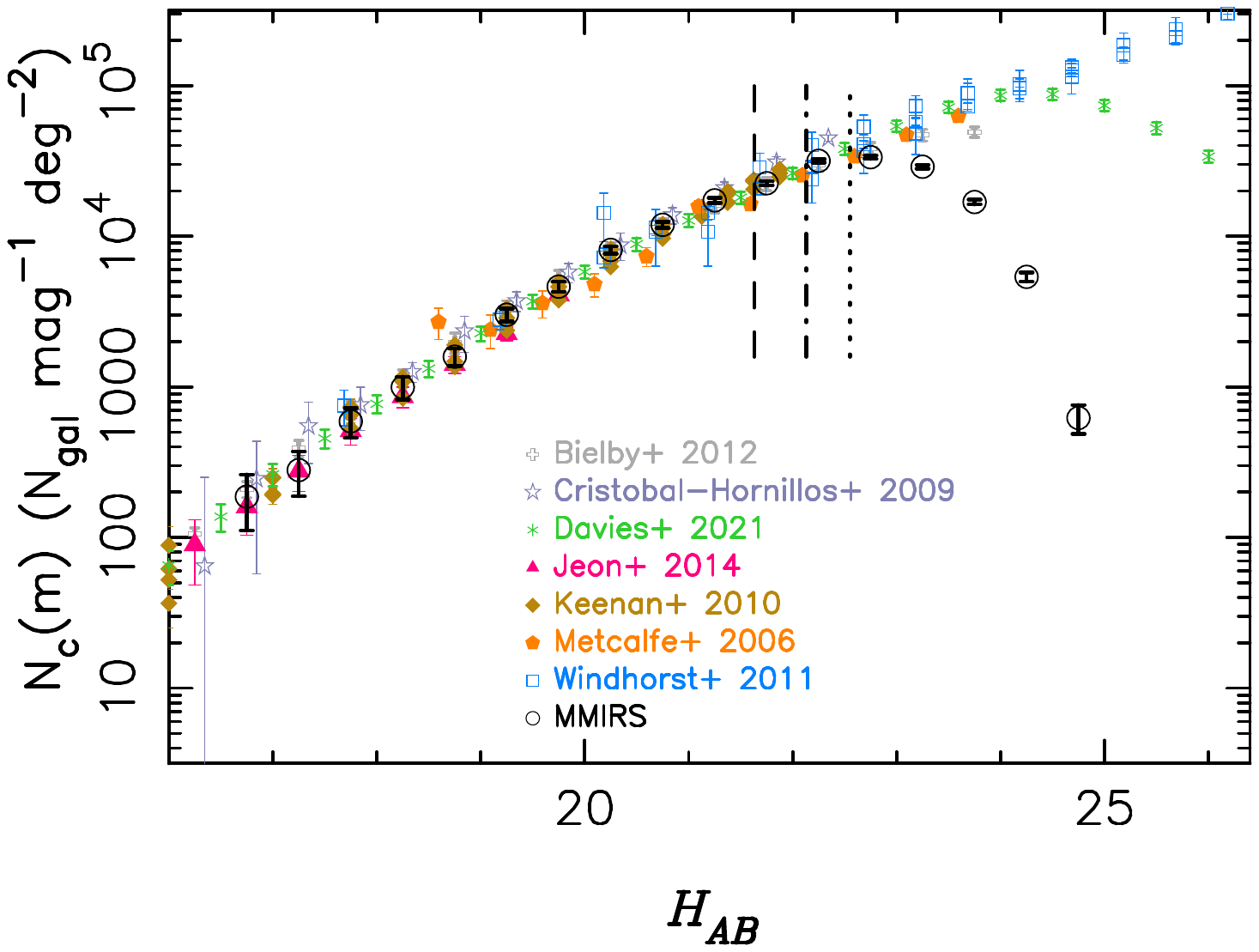}{0.45\textwidth}{(c)}}
{\fig{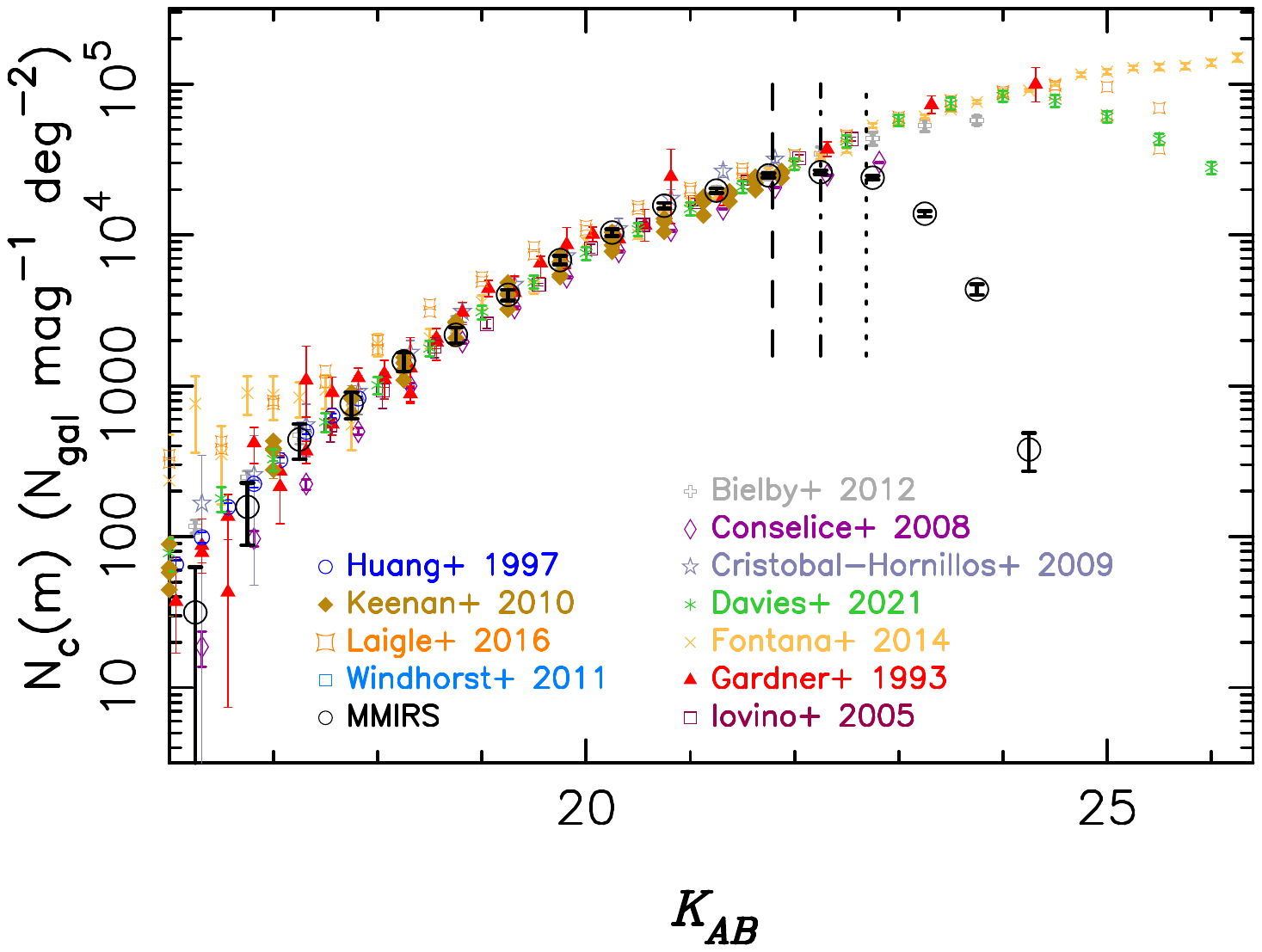}{0.45\textwidth}{(d)}}
}
\caption{Differential galaxy number counts ($N_c(m)$) measured in this
  work (black open circles) compared to other measurements in the
  literature (symbols as noted in the key) for the NIR filters. The
  vertical line segments indicate the location of the 95\%, 80\% and
  50\% completeness levels of Table~\ref{tab:completeness} for
  extended sources as dashed, dot-dashed and dotted lines
  respectively. Data being plotted come from \citet{Bielby2012},
  \citet{Conselice2008}, \citet{Cristobal-hornillos2009},
  \citet{Davies2021}, \citet{Fontana2014}, \citet{Gardner1993},
  \citet{Iovino2005}, \citet{Huang1997}, \citet{Jeon2014}, \citet{Keenan2010}, \citet{Laigle2016}, \citet{Metcalfe2006}, \citet{Windhorst2011}. For references where Table~\ref{tab:number_counts_altri} tabulates magnitude bins = 0.5, the number counts in galaxies 0.5 mag$^-1$ per unit solid angle were converted into  galaxies per unit magnitude per unit solid angle.\label{fig:number_counts}}
\end{figure*}
\begin{deluxetable*}{crrrrrrrr}[htb!]
\tabletypesize{\small}
\tablecaption{Number counts.\label{tab:number_counts_table}}\tablenotemark{a}
\tablewidth{0pt}
\tablehead{
\colhead{Filter} & 
\colhead{$N_c$(\textsl{Y})} & uncertainty &
\colhead{$N_c$(\textsl{J})} & uncertainty &
\colhead{$N_c($\textsl{H})} & uncertainty &
\colhead{$N_c$(\textsl{K})} & uncertainty\\
\colhead{ Magnitude} & 
\colhead{N/Mag/deg$^{2}$}& \colhead{N/Mag/deg$^{2}$} &
\colhead{N/Mag/deg$^{2}$}& \colhead{N/Mag/deg$^{2}$} &
\colhead{N/Mag/deg$^{2}$}& \colhead{N/Mag/deg$^{2}$} &
\colhead{N/Mag/deg$^{2}$}& \colhead{N/Mag/deg$^{2}$}}
\startdata
16.25 &     {...} &     {...} & {...} & {...} & {...} & {...} & 31.5 & 31.4\\
16.75 &     94.4 &     54.1 & 206.8 & 68.3 & 186.6 & 75.3 & 157.7 & 69.8\\
17.25 &    157.4 &     69.6 & 298.8 & 81.9 & 279.9 & 92.0 & 441.6 & 115.9\\
17.75 &    251.8 &     87.8 & 390.7 & 93.5 & 590.9 & 132.8 & 757.1 & 150.9\\
18.25 &    629.5 &    137.8 & 965.3 & 145.8 & 995.2 & 171.3 & 1451.1 & 206.9\\
18.75 &    912.8 &    165.1 & 1585.8 & 185.7 & 1586.1 & 214.6 & 2176.7 & 251.5\\
19.25 &   1857.1 &    232.9 & 2298.3 & 222.3 & 3016.8 & 292.1 & 4006.4 & 336.1\\
19.75 &   2486.6 &    267.9 & 3723.3 & 280.3 & 4634.0 & 357.9 & 6814.0 & 430.6\\
20.25 &   4375.2 &    350.2 & 7009.9 & 378.4 & 8117.3 & 464.3 & 10378.8 & 521.9\\
20.75 &   7648.7 &    454.1 & 9698.9 & 440.3 & 11911.6 & 552.7 & 15615.5 & 626.2\\
21.25 &  12118.3 &    559.7 & 15375.8 & 544.0 & 17323.1 & 652.6 & 19653.4 & 692.2\\
21.75 &  17909.9 &    665.3 & 20570.0 & 620.1 & 22485.8 & 730.6 & 24827.0 & 764.7\\
22.25 &  24268.1 &    758.1 & 28338.3 & 714.1 & {\it{31660.5}} & {\it{843.5}} & {\it{26057.3}} & {\it{780.4}}\\
22.75 & {\it{34529.3}} & {\it{877.6}} & {\it{33302.6}} & {\it{765.7}}& {\it{33495.4}} & {\it{863.2}} & {\it{23943.7}} & {\it{753.1}}\\
23.25 & {\it{38778.6}} & {\it{919.6}} & {\it{34681.6}} & {\it{779.1}} & {\it{28954.7}} & {\it{812.9}} & 1{\it{3817.3}} & {\it{593.3}}\\
23.75 &  {\it{35851.3}} & {\it{891.1}} & {\it{25074.7}} & {\it{676.9}} & {\it{16887.7}} & {\it{645.4}} & {\it{4384.9}} & {\it{350.7}}\\
24.75 &  {\it{ 7617.2}} & {\it{453.3}} & {\it{4458.7}} & {\it{305.5}} & {\it{622.0}} & {\it{136.2}} & {...} & {...}\\
\enddata
\tablenotetext{a}{Italics refer to counts affected by incompleteness estimated from the Monte Carlos simulations of Section 4.1.}
\end{deluxetable*}

In addition to characterizing the large-scale properties of a sample
(e.g., Koo \& Kron \citeyear{KooKron1992}, Smail et
al. \citeyear{Smail1995}), the number counts (shown in
Table~\ref{tab:number_counts_table}) provide an independent check on
the quality of the photometry, the effectiveness of the star/galaxy
separation (e.g., Conselice et al. \citeyear{Conselice2008}) and of the catalog completeness \citep[e.g.,][]{Windhorst2023}.  Figure~\ref{fig:number_counts} shows the number counts $N_c(m)  = \frac{N(m)}{\Delta m \times {\rm area}}$ measured in the \TDF\ for the \textsl{Y}, \textsl{J}, \textsl{H}, and the combined \textsl{K} sample. The uncertainties in Table~\ref{tab:number_counts_table} assume  Poisson statistics $\delta N_c(m) = (\sqrt{N(m)} \times {\rm area} - N(m) \times \Delta {\rm area})/(\Delta m \times {\rm area})^2$  and are underestimated at larger numbers of galaxies because no  cosmic variance is included.
Vertical lines representing the completeness levels for extended sources from Table~\ref{tab:completeness} are also shown and indicate that the turnover due to incompleteness is well characterized by the Monte Carlo simulations.
The figures also show a selection of published measurements noted in Table~\ref{tab:number_counts_altri}, and when required, the latter were converted into AB magnitudes by adding the Vega to AB offsets of (\textsl{J}, \textsl{H}, \textsl{K}) = (0.870, 1.344, 1.814) mag derived for the 2MASS \citep{Skrutskie2006} filters by \citet{2018ApJS..236...47W}. The published counts in $N_{gal}~0.5 mag^{-1}$  per unit  solid angle, where converted into $N_{gal}~mag^{-1}~deg^{-2}$. The areas and limiting magnitudes come from the faintest magnitudes in the original references\footnote{In the case of \citet{Gardner1993}, the authors also included earlier measurements which cover the range of areas noted in the table.}.
Given the small area covered by the \TDF\ MMIRS imaging---(0.0635, 0.0870, 0.0643, 0.0634) deg$^2$ for (\textsl{Y}, \textsl{J}, \textsl{H}, \textsl{K}) respectively)---some differences relative to other measurements are expected, though the agreement is good for magnitudes of 18 and fainter in all four bands. 

\begin{deluxetable*}{lcccc}
\tabletypesize{\small}
\tablecaption{Depth and area of literature number counts.\label{tab:number_counts_altri}}
\tablewidth{0pt}
\tablehead{
\colhead{Reference} & 
\colhead{Filter} & 
\colhead{magnitude bin} &
\colhead{Area} &
\colhead{Limiting magnitude [AB]\tablenotemark{a}}}
\startdata
\citet{Davies2021}    & \textsl{Y} & 0.5 & 0.6 deg$^2$     & 25.5\\ 
\citet{Windhorst2011} & \textsl{Y} & 0.5 & $\sim$ 45arcmin$^2$ & 26.0\\
{}                    &     {}     &   {}          &  {}  \\
\citet{Bielby2012}    & \textsl{J} & 0.5 & 2.1 deg$^2$ & 23.8 \\
\citet{Conselice2008} & \textsl{J} & 1.0 & 1.53 deg$^2$  & 23.4\\
\citet{Cristobal-hornillos2009}&  \textsl{J} & 1.0 & 0.5 deg$^2$ &23.3 \\
\citet{Davies2021}    & \textsl{J} & 0.5 & 6 deg$^2$      & 25.2\\ 
\citet{Iovino2005}    & \textsl{J} & 0.5 & 400 arcmin$^2$ & 22.4\\
\citet{Jeon2014}      & \textsl{J} & 0.5 & 5.1 deg$^2$  &  20.75\\
\citet{Keenan2010}    & \textsl{J} & 1.0 & 2.77 deg$^2$   & 23.0 to 24.4\\
\citet{Windhorst2011} & \textsl{J} & 0.5 & $\sim$ 45 arcmin$^2$ & 26.0\\
{} &{} &{} &{}\\
\citet{Bielby2012}    & \textsl{H} & 0.5 & 2.1 deg$^2$ & 23.8\\
\citet{Cristobal-hornillos2009}&  \textsl{H} & 1.0 & 0.5 deg$^2$ &22.6 \\
\citet{Davies2021}    & \textsl{H} & 0.5 & 6 deg$^2$     & 25.0\\ 
\citet{Jeon2014}      & \textsl{H} & 0.5 & 5.1 deg$^2$   & 20.75\\
\citet{Keenan2010}    & \textsl{H} & 1.0 & 3.6 deg$^2$   & 22.5 to 24.0\\
\citet{Metcalfe2006}  & \textsl{H} & 0.5 & 49 arcmin$^2$ & 22.5\\
\citet{Windhorst2011} & \textsl{H} & 0.5 & $\sim$ 45 arcmin$^2$ & 26.0\\
{}                    &     {}     &   {}          &  {}  \\
\citet{Bielby2012}    & \textsl{K} & 0.5 & 2.1 deg$^2$  & 23.8\\
\citet{Conselice2008} & \textsl{K} & 1.0 & 1.53 deg$^2$  & 23.8\\
\citet{Cristobal-hornillos2009}& \textsl{K} & 1.0 & 0.5 deg$^2$ &21.8 \\
\citet{Davies2021}    & \textsl{K} & 0.5 &6 deg$^2$     &24.8 \\
\citet{Fontana2014}   & \textsl{K} & 1.0 & 340.58 arcmin$^2$ & 26.5\\
\citet{Gardner1993}   & \textsl{K} & 0.5 & 1.16 arcmin$^2$ to 41.56 deg$^2$ & 24.3\\
\citet{Huang1997}     & \textsl{K} & 1.0 & 1.58 deg$^2$   & 16.3\\
\citet{Iovino2005}    & \textsl{K} & 0.5 & 400 arcmin$^2$ & 22.6\\
\citet{Keenan2010}    & \textsl{K} & 1.0 & 3.94 deg$^2$  & 23.0 to 24.8\\
\citet{Laigle2016}    & \textsl{K} & 0.5 & 0.62 deg$^2$  & 24.7 \\
\enddata
\tablenotetext{a}{Values quoted by the original references}
\end{deluxetable*}

The deepest data in \textsl{Y, J}, and \textsl{H} come from the HST imaging in GOODS-S by the Early Science Release using the Wide-Field Camera 3 of \citet{Windhorst2011}. In \textsl{K}, the deepest coverage comes from the ground-based observations of \citet{Fontana2014} in the UDS and GOODS-S fields. 
Because of the large number of samples in \textsl{K}, it is also the band showing the greatest variation between samples. The greatest differences are seen for the \textsl{K} number counts at bright magnitudes for the \citet{Bielby2012} and \citet{Fontana2014} because of small-number statistics. Fainter than \textsl{K} $\sim$ 18, the largest differences are seen for the \citet{Iovino2005} and \citet{Conselice2008} counts.
However, the \textsl{J} counts of both references are good agreement with other works. For \textsl{J}, \textsl{H} and \textsl{K} the MMIRS counts show good agreement with the multi-field number counts of \citet{Keenan2010} throughout the common magnitude range.


\section{Conclusion}
We present the results of near infrared imaging in the JWST North Ecliptic
Pole Time Domain Field in the \textsl{Y},  \textsl{J}, \textsl{H} and
\textsl{K} bands. These observations reach a 95\% completeness level for AB magnitudes of
(\textsl{Y, J, H, K})  = (22.3, 21.8, 21.6, 21.8) estimated from simulations of extended sources described by an exponential S\'ersic profile. For point sources, the corresponding 95\% completeness levels are (\textsl{Y, J, H, K})  = (23.8, 23.5, 23.1, 23.3) in AB magnitudes. 
These data are combined with catalogs derived from archival Subaru Hyper-Suprime-Cam observations in the (\textsl{g}, \textsl{i2}, \textsl{z}, \textsl{NB816}, \textsl{NB921}) bands to create a matched visible--NIR catalog, with positional precision with a dispersion $\sim$ 20 milli-arcseconds.   A comparison with a catalog of sources from the first two JWST observations  of the \TDF\ suggests that a small number of HSC-MMIRS sources are galaxies at redshifts $z\gtrsim$~1.5 and above or dusty galaxies. The expected number of galaxies at redshifts $\gtrsim$  2 is estimated to be $\sim$ 10 using the redshift versus number density diagram \citet{Conselice2008} (their Figure 3), who reach a comparable depth in the \textsl{K} band imaging, though this number is very uncertain, particularly given the small area the MMIRS imaging covers.
The examination of bright unsaturated sources in the joint HSC-MMIRS-JWST shows no source with significant ($\Delta$m $\geq$ 0.1 mag) variability.
These observations provide the first epoch deep NIR photometry in this part of the sky and will continue being used in the search for transient and variable objects in the \TDF. This region of the sky has already amassed in very few years an extensive coverage in wavelength space, ranging from X-rays to radio frequencies.

The mosaics and catalogs are publicly available at
{\url{https://sites.google.com/view/jwstpearls/data-products}} and on Zenodo under an open-source 
Creative Commons Attribution license:
\dataset[doi:10.5281/zenodo.7934393]{https://doi.org/10.5281/zenodo.7934393}

We would like to dedicate this paper to the memory of our PEARLS colleague Mario Nonino and acknowledge his suggestions in preparing images, catalogs and documentation for the public release.
We respectfully acknowledge that the University of Arizona and Arizona State University are on the land and territories of Indigenous peoples. Today, Arizona is home to 22 federally recognized tribes, with Tucson being home to the O'odham and the Yaqui and Tempe home to the Akimel O’odham
(Pima) and Pee Posh (Maricopa). Committed to diversity and inclusion, the University of Arizona strives to build sustainable relationships with sovereign Native Nations and Indigenous communities through education offerings, partnerships, and community service.
We wish to recognize and acknowledge the very significant cultural role and reverence that the summit of
Mauna Kea has always had within the Kānaka Maoli community. We are most fortunate to have the opportunity to conduct observations from this mountain.
We thank Brian McLeod (MMIRS PI) and Igor Chilingarian for their
assistance and insight in preparing and reducing the MMIRS data and
for providing the filter throughputs used in the photometric
calibration. We also thank Dallan Porter and Sean Moran
for their assistance in implementing more MMTO data retrieval options and Catherine Grier for proving a script that enabled calculating exposure maps using \texttt{swarp}. We thank the anonymous referee for comments and suggestions that allowed improving the presentation.
CNAW and MJR acknowledge support from the NIRCam Development Contract NAS5-02105 from NASA Goddard Space Flight Center to the University of Arizona.
CNAW and RAJ gratefully acknowledge funding from the HST-GO-15278.008-A grant. 
RAW, SHC, and RAJ acknowledge support from NASA JWST Interdisciplinary Scientist grants NAG5-12460, NNX14AN10G and 80NSSC18K0200 from GSFC.
JFB was supported by NSF grant No. PHY-2012955. 
IS acknowledges support from STFC (ST/T000244/1).
MI acknowledges the support from the National Research Foundation of Korea (NRF) grants, No. 2020R1A2C3011091, and No. 2021M3F7A1084525, funded by the Korean government (MSIT). M.H. acknowledges the support from the Korea Astronomy and Space Science Institute grant funded by the Korean government (MSIT) (No. 2022183005) and the support from the Global Ph.D. Fellowship Program through the National Research Foundation of Korea (NRF) funded by the Ministry of Education (NRF-2013H1A2A1033110).
We acknowledge the SMOKA archive of the National Astronomical Observatory of Japan from which Subaru Hyper-Suprime-Cam data were retrieved. This research has made use of the NASA/IPAC Extragalactic Database, which is funded by the National Aeronautics and Space Administration and operated by the California Institute of Technology.
This work has made use of data from the European Space Agency (ESA) mission
{\it Gaia} (\url{https://www.cosmos.esa.int/gaia}), processed by the {\it Gaia}
Data Processing and Analysis Consortium (DPAC,
\url{https://www.cosmos.esa.int/web/gaia/dpac/consortium}). Funding for the DPAC
has been provided by national institutions, in particular the institutions
participating in the {\it Gaia} Multilateral Agreement.

\vspace{5mm}
\facilities{MMT (MMIRS), GAIA, HST (ACS, WFC3), NAOJ (Subaru \textsl{Hyper Suprime-Cam}, SMOKA), IRSA (2MASS), NED, MAST (Pan-STARRS), SAO/NASA (ADS), SDSS}.

\software{Fortran, Perl, IDL, 
  Astrometry.net ({\url{https://astrometry.net/}}, \citet{Lang2010}), 
  GalSim \citep{Rowe2015},  
  HSC pipeline \citep{Bosch2018},
  IDL Astronomy User's Library \citep{Landsman1995},
  IDL wrapper scripts for MMIRS reduction \citep{Willmer2023},
  Matplotlib \citep{Matplotlib},
  MMIRS pipeline \citep{Chilingarian2015},
  MPFIT \citep{Markwardt2009},
  PGPLOT ({\url{https://sites.astro.caltech.edu/~tjp/pgplot}}), 
  PSFEX \citep{Bertin2011}, SAOimage DS9 \citep{ds9},
  SCAMP \citep{Bertin2006}, SExtractor \citep{BertinArnouts},
  SWARP \citep{Bertin2002}, TOPCAT \citep{2005ASPC..347...29T}}.

\appendix
\section{MMIRS Observation log}
Table~\ref{tab:log} is the MMIRS observation log, which identifies the date of observation (UT), the modified Julian Date, the mean seeing value and dispersion measured from all images taken during the night, the
quadrant(s) and filter(s), and in parenthesis the gain values. When no
gain value is shown, the default 2.68 e$^-$/DN gain was used.
\begin{deluxetable*}{lcccl}[h]
\centering 
\tablecaption{MMIRS Observation log.\label{tab:log}}
\tablewidth{0pt}
\tablehead{
\colhead{date} & \colhead{MJD} & \colhead{Seeing} & \colhead{uncertainty} &
\colhead{field, filter}\\
\colhead{}     & \colhead{}     & \colhead{arcsec}     & \colhead{arcsec}     & \colhead{} 
}
\startdata
2017-05-14 & 57887 & 1.117  & 0.155  &   SE \textsl{J}, SW \textsl{J}\\
2017-06-11 & 57915 & 1.085  & 0.183  &   SW \textsl{J}, NE \textsl{J}\\
2017-06-12 & 57916 & 1.094  & 0.228  &   NW \textsl{J}\\
2017-06-13 & 57917 & 0.843  & 0.503  &   SE \textsl{K$_{\rm spec}$}\\
2017-06-14 & 57918 & 1.015  & 0.324  &   SE \textsl{K$_{\rm spec}$}\\
2017-06-15 & 57919 & 0.871  & 0.354  &   SE \textsl{K$_{0.95}$}, SW \textsl{K$_{0.95}$}, \textsl{K$_{2.68}$}\\
2017-06-16 & 57920 & 1.293  & 0.507  &   SW \textsl{K$_{2.68}$}\\
2017-06-18 & 57922 & 1.994  & 0.193  &   NE \textsl{K$_{2.68}$}\\
2017-06-19 & 57923 & 1.942  & 0.667  &   NE \textsl{K$_{2.68}$}\\
2017-07-01 & 57935 & 0.769  & 0.234  &   NE \textsl{K$_{2.68}$}\\
2017-07-02 & 57936 & 0.605  & 0.093  &   NE \textsl{K$_{2.68}$}\\
2017-07-06 & 57940 & 1.856  & 0.430  &   NE \textsl{K$_{2.68}$}\\
2017-10-17 & 58043 & 1.433  & 0.445  &   NE \textsl{K$_{2.68}$}, NW \textsl{K$_{2.68}$}\\
2017-10-19 & 58045 & 1.183  & 0.228  &   NW \textsl{K$_{2.68}$}\\ 
2017-10-20 & 58046 & 0.745  & 0.126  &   NW \textsl{K$_{2.68}$} \\
2017-11-03 & 58060 & 0.880  & 0.083  &   NW \textsl{K$_{2.68}$} \\
2017-11-07 & 58064 & 0.878  & 0.244  &   NW \textsl{K$_{2.68}$} \\
2018-08-28 & 58358 & 0.855  & 0.089  &   SE \textsl{H}, SE \textsl{Y}, NE \textsl{Y}\\
2018-08-29 & 58359 & 0.811  & 0.138  &   SW \textsl{Y}, NW \textsl{Y} \\
2018-08-31 & 58361 & 0.837  & 0.123  &   SE \textsl{H} \\
2018-11-17 & 58439 & 0.926  & 0.166  &   NW \textsl{H}\\
2018-11-19 & 58441 & 0.708  & 0.091  &   NW \textsl{H}\\
2018-11-20 & 58442 & 0.999  & 0.189  &   NW \textsl{H}\\
2018-11-22 & 58444 & 1.165  & 0.371  &   NW \textsl{H}\\
2018-11-25 & 58447 & 0.966  & 0.144  &   NW \textsl{H}, NE \textsl{H}\\
2018-11-28 & 58450 & 1.026  & 0.227  &   NE \textsl{H} \\
2019-06-14 & 58648 & 1.154  & 0.191  &   NE \textsl{H}\\
2019-06-15 & 58649 & 0.732  & 0.097  &   SW \textsl{H}\\
2019-06-16 & 58650 & 0.584  & 0.051  &   Center \textsl{J}\\
2019-06-18 & 58652 & 0.770  & 0.049  &   SW1 \textsl{J}\\
2019-06-21 & 58655 & 1.042  & 0.150  &   NW1 \textsl{J}\\
\enddata
\end{deluxetable*}

\begin{figure}[htb!]
\centering
\epsscale{0.70}
\plotone{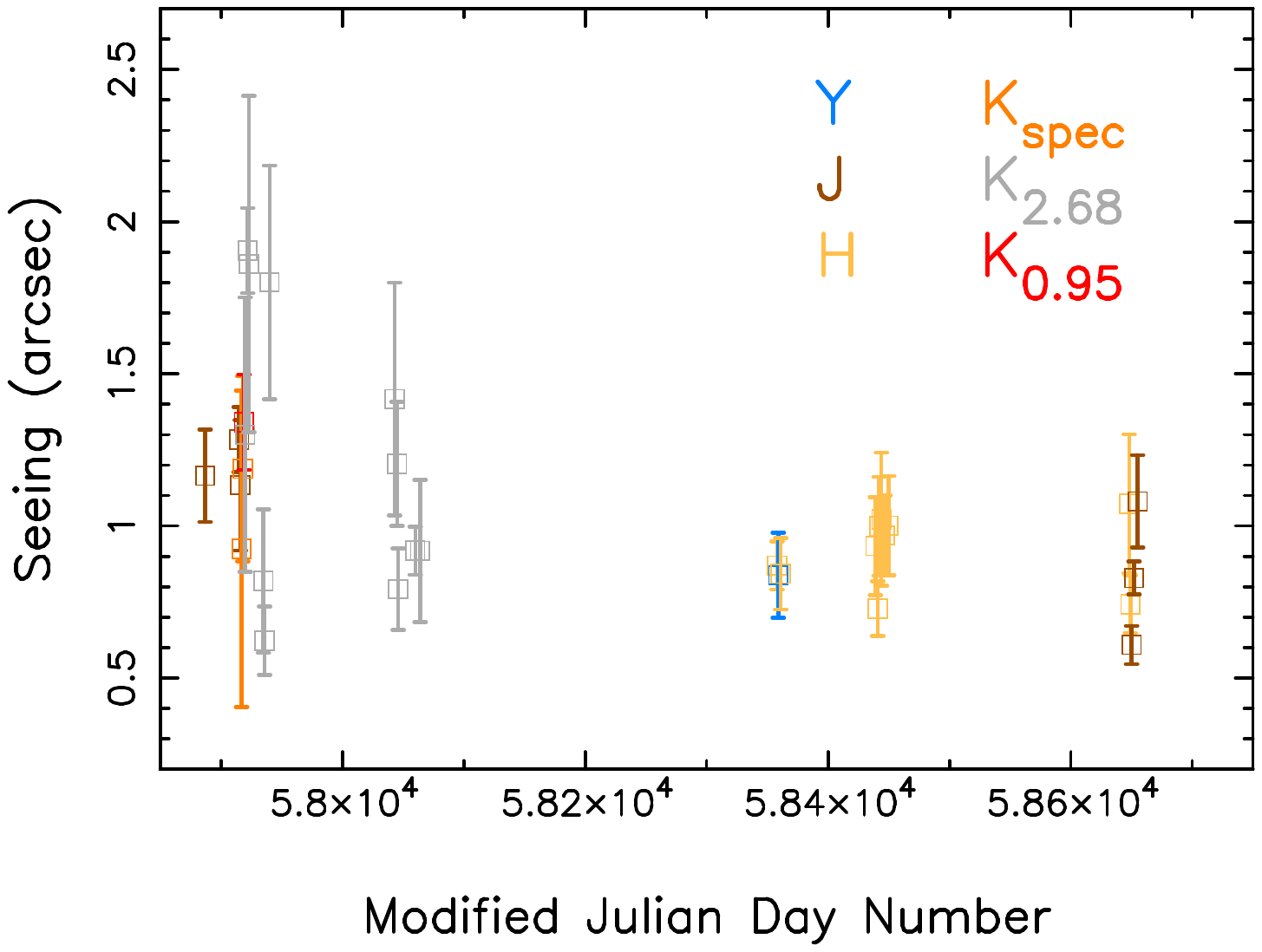}
\caption{Distribution of the average seeing FWHM and dispersion measured from all images during a night presented in Table~\ref{tab:log}. The colors represent different filters as noted in the key.\label{fig:seeinglog}}
\end{figure}

\section{\texttt{IDL} imaging wrapper}
A wrapper to run the initial imaging reduction scripts is available at
{\url{https://github.com/cnaw/mmirs\_imaging}} (also available at \citet{Willmer2023}),
though currently there is limited documentation on how to use them.
Most of the procedures are those of \citet{Chilingarian2015} but with
small modifications in the initial reduction
that converts data cubes into slopes. Also in the repository is a
\texttt{perl} script, \textbf{pre\_reduction.pl}, that
will read the contents of the directory where raw images are stored,
sort images according to the \textsf{DARKTIME} keyword value, and
check whether they are science or calibration data.
This script attempts to emulate the \texttt{perl} script described in
the \citet{Chilingarian2015} paper.
The pre-reduction script sorts imaging from spectroscopic data (this feature was not tested extensively) and creates a batch file that should be
run under \texttt{IDL}. 
The final product of this step is a set of dark-subtracted,
linearity-corrected slope images.





\end{document}